\documentclass[fleqn,usenatbib]{mnras}

\usepackage{newtxtext,newtxmath}

\usepackage{graphicx}	
\usepackage{amsmath}	
\usepackage[FIGTOPCAP]{subfigure}
\usepackage{caption}
\usepackage{bm}
\usepackage{soul}
\usepackage{multirow}
\usepackage[capitalise]{cleveref}

\title[Formation of moonlets confining arcs]{Numerical analysis of processes for the formation of moonlets confining the arcs of Neptune}

\author[Madeira \& Giuliatti Winter]
{Gustavo {Madeira},$^{1,2}$\thanks{E-mail: madeira@ipgp.fr} and
Silvia M. {Giuliatti Winter}$^{1}$\thanks{E-mail: giuliatti.winter@unesp.br}
\\
$^{1}$Grupo de Din\^amica Orbital \& Planetologia, S\~ao Paulo State University -UNESP, Av. Ariberto Pereira da Cunha, 333, Guaratinguet\'a SP, 12516-410, Brazil\\
$^{2}$Universit\'e de Paris, Institut de Physique du Globe de Paris, CNRS, F-75005 Paris, France}

\date{Accepted 2022 April 1. Received 2022 March 31; in original form 2021 August 25}

\pubyear{2021}

\begin{document}
\label{firstpage}
\pagerange{\pageref{firstpage}--\pageref{lastpage}}
\maketitle

\begin{abstract}
The arcs of Neptune - Fraternit\'e, Egalit\'e, Libert\'e, and Courage - are four incomplete rings immersed in the Adams ring. A recent confinement model for the arcs proposes that the structures are azimuthally confined by four co-orbital moonlets. In this work, we intend to approach some points related to the dynamics of co-orbital moonlets and suggest a model for their formation. We study the equilibrium configurations for 1+$N$ co-orbital satellites under the 42:43 Lindblad resonance with Galatea. We obtained three distinct configurations with 1+3 and 1+4 moonlets able to confine and reproduce the location of the arcs. The moonlets' formation is analysed by the disruption of an ancient body at a Lagrangian point of a moon. The disruption fragments spread out in horseshoe orbits and collide to form  moonlets, which reach an equilibrium configuration due to a non-conservative effect. In such a scenario, the arcs likely formed through a mixture of different processes, with impacts between disruption outcomes and meteoroid impacts with the moonlets being possibilities.
\end{abstract}

\begin{keywords}
planets and satellites: dynamical evolution and stability -- planets and satellites:
rings -- planets and satellites: formation -- celestial mechanics
\end{keywords}
%
\section{Introduction}
In 1984, during a stellar occultation, an incomplete ring was detected around the planet Neptune \citep{Hu86}. Confirmed by ground-based observations \citep{Si91} and \textit{Voyager} spacecraft images \citep{Sm89}, the four arcs of Neptune, known as Fraternit\'e, Egalit\'e, Libert\'e, and Courage, are indeed the densest parts of a complete ring, the Adams ring. They have individual angular widths ranging from 2~deg (Courage) to 9~deg (Fraternit\'e) and radial width of 15~km \citep{Po95}. Since differential Keplerian motion would completely spread the arcs in about three years \citep{Pa19}, several confinement models were proposed over time to explain these structures' dimensions and stability.

The first known confinement model is present in \cite{Br11}, in which the author shows that Jupiter confines trojan asteroids in tadpole orbits around its triangular points ($L_4$ and $L_5$) in the Sun-Jupiter-trojans system. Based on this work, \cite{Li85} proposed that a sizeable hypothetical satellite would azimuthally confine the Neptune arcs in its triangular points. At the same time, another hypothetical internal satellite would be responsible for the radial confinement of the arcs. \cite{Si92} improved such model by proposing that a pair of co-orbital satellites azimuthally confine the structures, which allows the possible existence of smaller satellites. However, such models were ruled out since \textit{Voyager} spacecraft did not observe satellites with the dimensions required by them.

A confinement model envisioned by \cite{Go86} and confirmed by the discovery of the satellite Galatea \citep{Sm89} proposes that a single internal satellite would be responsible for the azimuthal and radial confinements of the arcs. \cite{Po91} shows that the arcs are close to the 84:86 corotation inclined resonance (CIR, azimuthal confinement) and the 42:43 Lindblad resonance (LER, radial confinement) with Galatea. The author proposes that the arcs are trapped in some of the 84 sites formed by the CIR, which could explain their radial and azimuthal widths \citep{Fo96}. Similarly, the coupling between Lindblad and corotation resonances with Mimas is the mechanism that holds (at least, temporarily) the Aegaeon, Anthe, and Methone arcs of Saturn \citep{He09,He10,Mo13,Su17,Ma18,Ma20}.

New evidence from ground-based observations shows that the semi-major axis of the arcs is displaced from the 84:86 CIR location \citep{Si99,Du02}, which leaves the arcs without azimuthal confinement. The arcs have changed location and decayed in intensity since their discovery \citep{Pa05,Sh13,Re14}. In fact, data discussed in \cite{Pa19} indicate the disappearance of the arcs Libert\'e and Courage.

\cite{Re14} rescue the confinement model based on co-orbital satellites, proposing that Galatea radially confines the arcs while several co-orbital moonlets (at least four) with diameters below the precision of \textit{Voyager} spacecraft confine them azimuthally. The system was assumed to consist of four co-orbital moonlets and Galatea. A set of azimuthal locations and mass ratios is obtained for the co-orbital satellites to reproduce the arc widths. Next, a representative case is explored by the authors where the masses of the moonlets $S_1$, $S_2$, $S_3$ and $S_4$, are assumed to be $60.0$, $0.54$, $1.17$ and $0.66 \times 10^{13}$~kg, respectively. $S_2$, $S_3$ and $S_4$ are azimuthally located at longitudes ${\rm \theta=48.31}$~deg, ${\rm 59.38}$~deg and ${\rm 72.19}$~deg, respectively, with respect to $S_1$ (${\rm \theta=\lambda-\lambda_{S_1}}$, where $\lambda$ is the mean longitude). All the system is in the 42:43 LER with Galatea.

\cite{Gi20} explore the model proposed in \cite{Re14}, including the effects of the solar radiation force and also accounting for the mass production rate of the moonlets. They found that micrometre particles are removed from the arcs due to solar radiation in less than 50~years. Part of these particles become transient between the arcs, being a possible explanation for the arcs' changes in longitude and intensity. In this work, we intend to approach some points related to the dynamics of these co-orbital satellites and propose a model for their formation.

The 1:1 resonance dynamics have been known for over a century in the planar restrict 3-body problem. It has been used to explain the dynamics of the trojan asteroids \citep{Br11}, Helene and Polydeuces \citep{Le80,Po07}, -- confined in the $L_4$ and $L_5$ points of Dione, respectively -- and Telesto and Calypso \citep{Sm80,Pa80,Ob03} -- confined in the $L_4$ and $L_5$ points of Tethys, respectively. Another example is the Janus and Epimetheus system, in which both satellites have comparable masses and perform horseshoe fashion orbits in the rotating frame, as shown by \cite{De81} and \cite{Yo83}.

The study of systems with more than two co-orbital satellites became feasible with computational advances. \cite{Sa88} carried out complete analytical and numerical studies of $N$ co-orbital satellites with the same mass ($N\leq9$) in circular orbits, identifying stable equilibrium configurations. Similar work was carried out by \cite{Re04}, where the authors solved analytically the case N=3 and proposed a numerical method to find the possible linearly stable solutions for any given set of masses and number of co-orbital satellites. More recently, \cite{He21} analysed the confinement of four D68 clumps by co-orbital satellites, obtaining a set of five moonlets capable of confining them. However, the authors rule out this scenario as a likely explanation for the clumps, given the highly fragile stability shown by the co-orbital configuration.

Here, we based our analysis on the work of \cite{Re04} to obtain the equilibrium configurations of 1+$N$ co-orbital satellites. We evaluate the effects of Galatea in the co-orbital system and also propose different scenarios for the formation of the arcs. \cite{Tr15} showed that the collisions of large fragments produced in the disruption of an ancient satellite might form Janus and Epimetheus. Following this work, we suggest the formation of co-orbital satellites through the disruption of an ancient body located at a triangular point of a satellite.

In Section~\ref{dyn}, we obtain the equilibrium positions of moonlets and particles in an eccentric system of 1+$N$ co-orbital satellites. The effects caused by  Galatea are analysed in Section~\ref{lersys}. In Section~\ref{4frag}, we study the temporal evolution of fragments from the disruption of an ancient body. In Section~\ref{arcsloc}, we present the discussion on the implications of our model for the formation of the arcs. A general discussion is addressed in Section~\ref{discussion}, and we present our conclusions in Section~\ref{conclusions}.

\section{1+$N$ co-orbital satellite dynamics} \label{dyn}

In this section, we revisit the work of \cite{Re04} and investigate the dynamics of a system with a gravitationally dominant satellite sharing its orbit with $N$ smaller satellites and a set of particles. We will assume the  dominant satellite as the satellite ${\rm S_1}$ proposed in \cite{Re14} in an eccentric orbit ($e=3\times 10^{-4}$, Table~\ref{initial}). All results will be given in the rotating frame with ${\rm S_1}$.

For clarity, we will refer to the largest co-orbital as ``moon'' and the $N$ smaller ones as ``moonlets'', keeping the ``satellite'' nomenclature for Galatea. Moonlets and particles differ from each other by the fact that the moonlets interact gravitationally with each other. In contrast, the particles feel the gravitational effect of the massive bodies but do not interact with each other. Consequently, each of these classes will have different stable equilibrium positions. Firstly, we analyse the equilibrium positions of the moonlets.  

\subsection{Moonlet stable equilibrium positions} \label{sqm}
\begin{figure}
\centering
\includegraphics[width=\columnwidth]{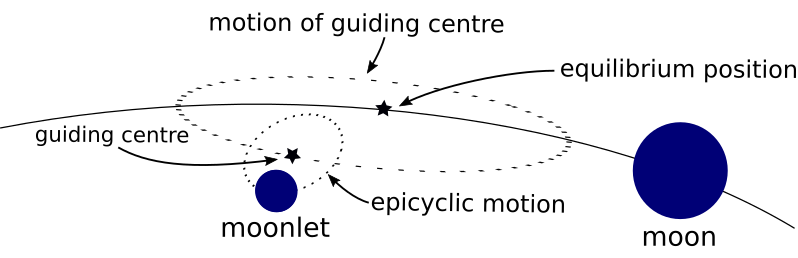}
\caption{Scheme of the trajectory in the rotating frame of a moonlet co-orbital to a larger  moon. Based on figure~3.14 of \citetalias{MD99}.}
\label{scheme}
\end{figure}

Assuming a system composed only by the planet and the 1+$N$ co-orbital satellites, we obtain tadpole-like trajectories for the $N$ moonlets in the rotating frame. Such trajectories are composed of two distinct motions: an epicyclic motion and the guiding centre (or epicycle centre) motion \citep{De84}. Figure~\ref{scheme} shows a scheme with the trajectory of a moonlet in the rotating frame. The equilibrium positions are locations of maximum potential energy and depend on the mass distribution in the system. When all moonlets are precisely in the equilibrium positions, they remain stationary in relation to each other, and we say that the system is in an equilibrium configuration.

\defcitealias{MD99}{Murray \& Dermott (1999)}

\begin{table*}
\centering
\begin{tabular}{llllrrrr}
\hline\noalign{\smallskip}
& $a$ (km) & $e$ ($10^{-4}$) & $I$ (deg) & $\varpi$ (deg) & $\Omega$ (deg) & $\lambda$ (deg) &  $m$ (kg)\\
\hline\noalign{\smallskip}
Galatea & 61953.0 & 2.2 & 0.0231 & 225.81 & 196.94 &  351.114 & 1.94 $\times$ 10$^{18}$ \\
$S_1$ &  62932.7 & 3.0 & 0.0 & 50.82 & 0.0 & 211.88 & 6.00 $\times$ 10$^{14}$ \\
\hline\noalign{\smallskip}
\end{tabular}
\caption{Orbital elements and masses of Galatea and the hypothetical moon $S_1$ \citep{Gi20}.}
\label{initial}
\end{table*}

We obtain the equilibrium configurations of 1+$N$ co-orbital satellites by performing numerical simulations with the \emph{Mercury} package, with the Bulirsch–Stoer algorithm \citep{Ch99}. We include Neptune with its gravitational coefficients ($J_2$ and $J_4$, the planet parameters were taken from \cite{Ow91}), the moon $S_1$ and $N$ moonlets. The initial orbital elements of $S_1$ and Galatea (later added to the system) are the same as those used in \cite{Gi20}, given in Table~\ref{initial}. Such values were obtained by \citeauthor{Gi20} to reproduce the observational data \citep{Po91} and the results of \cite{Re14}.

We have assumed moonlets with masses $m=10^{-2}m_{S_1}$, where $m_{S_1}$ is the mass of $S_1$, in the same orbit as the moon, but with different and randomly selected mean longitudes. We have also included a non-conservative term in the velocities to vary the system energy and carry the moonlets to the linearly stable equilibrium points \citep{Re04}. The term, provided in \cite{Re04} for circular orbits, is given by:
\begin{equation}
\dot{r}=-\nu(r-r_0) \label{dotr}
\end{equation}
where $r$ is the orbital radius of the body, $\dot{r}$ its temporal derivative, $r_0$ is the average orbital radius and $\nu$ is a constant that defines the timespan for the system to reach an equilibrium configuration.

\begin{figure}
\subfigure[$\nu=0$]{\includegraphics[width=\columnwidth]{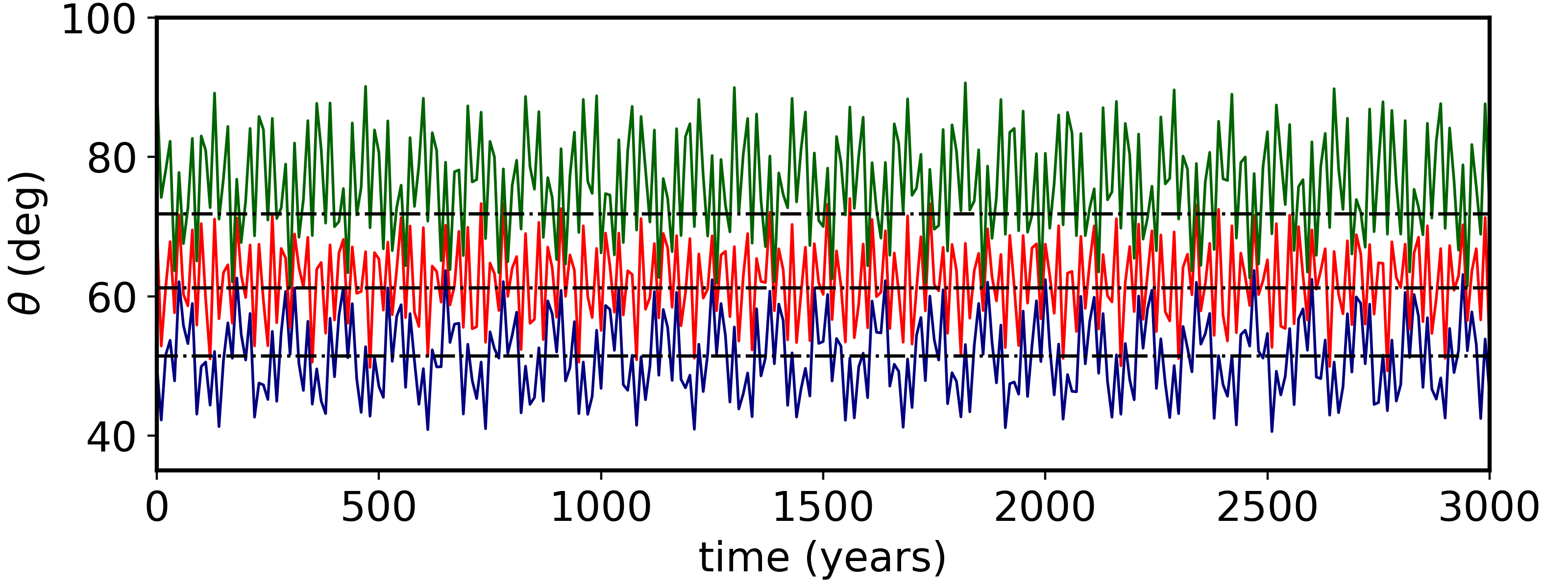} \label{woutd}}
\subfigure[$\nu=10^{-6}$~yrs$^{-1}$]{\includegraphics[width=\columnwidth]{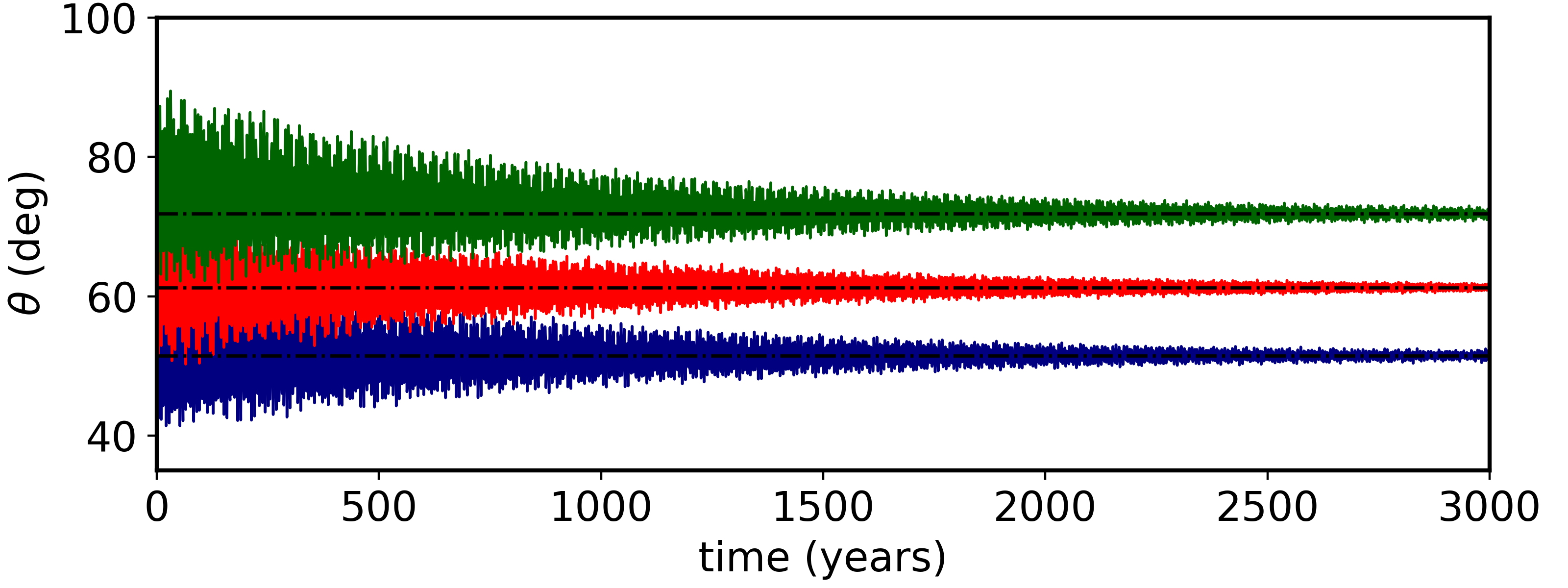} \label{withd}}
\caption{Azimuthal angle ($\theta=\lambda-\lambda_{S_1}$) of $S_1$ and three test moonlets initially at $\theta$=50~deg (solid blue line), 70~deg (solid red line) and 90~deg (solid green line) for a) $\nu=0$ and b) $\nu=10^{-6}$~yr$^{-1}$. The dotted lines correspond to the equilibrium position associated with each moonlet.}
\label{term}
\end{figure}

To demonstrate the effect of this non-conservative term, we present in Figure~\ref{term} the angular evolution of three moonlets initially at $\theta=50$, $70$, and $90$~deg, for $\nu=0$ and $10^{-6}$~yr$^{-1}$, from top to bottom. The moonlets are initially close to the equilibrium positions at $\theta=51.5$, $61.3$, and $71.9$~deg (vertical dotted lines). As we can see, without the non-conservative term (top panel), the moonlets remain in a significant angular motion around the equilibrium positions of the system. Meanwhile, when we include the non-conservative term (bottom panel), the amplitude of motion of the guiding centre decreases with time, and the moonlets spiral toward the equilibrium positions \citep{Re04}. The system reaches the equilibrium configuration in approximately $3000$~years. The rate at which the amplitude of motion decreases depends linearly on the value of $\nu$. For example, for $\nu=~10^{-5}$ and $10^{-4}$~yr$^{-1}$, the system reaches the equilibrium configuration in $300$, and $30$~years, respectively.
\begin{figure}
\centering
\includegraphics[width=\columnwidth]{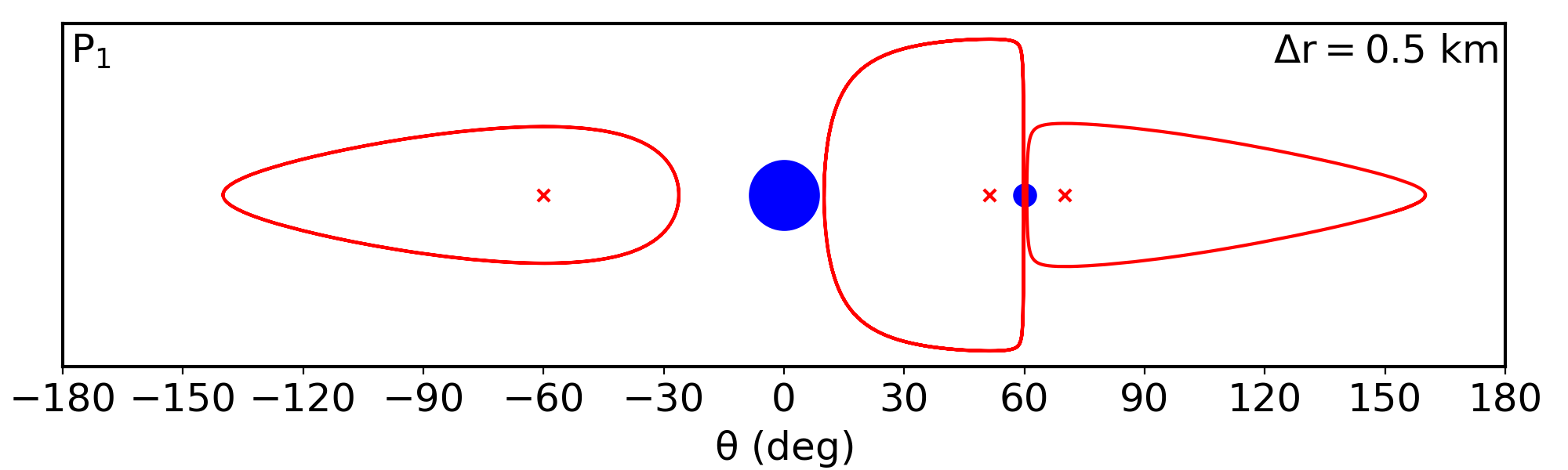}
\caption{Equilibrium positions of moonlets (small blue dots) and equilibrium positions of massless particles (red crosses) in a 1+1 co-orbital satellite system. The $x$-axis gives the longitude $\theta$ in relation to the moon (largest blue dot), and the $y$-axis shows the radial variation with scale $\Delta r$ given in the upper right corner of the figure. The red lines show the trajectory of some representative particles.}
\label{ptos2p}
\end{figure}
\begin{figure}
\centering
\includegraphics[width=\columnwidth]{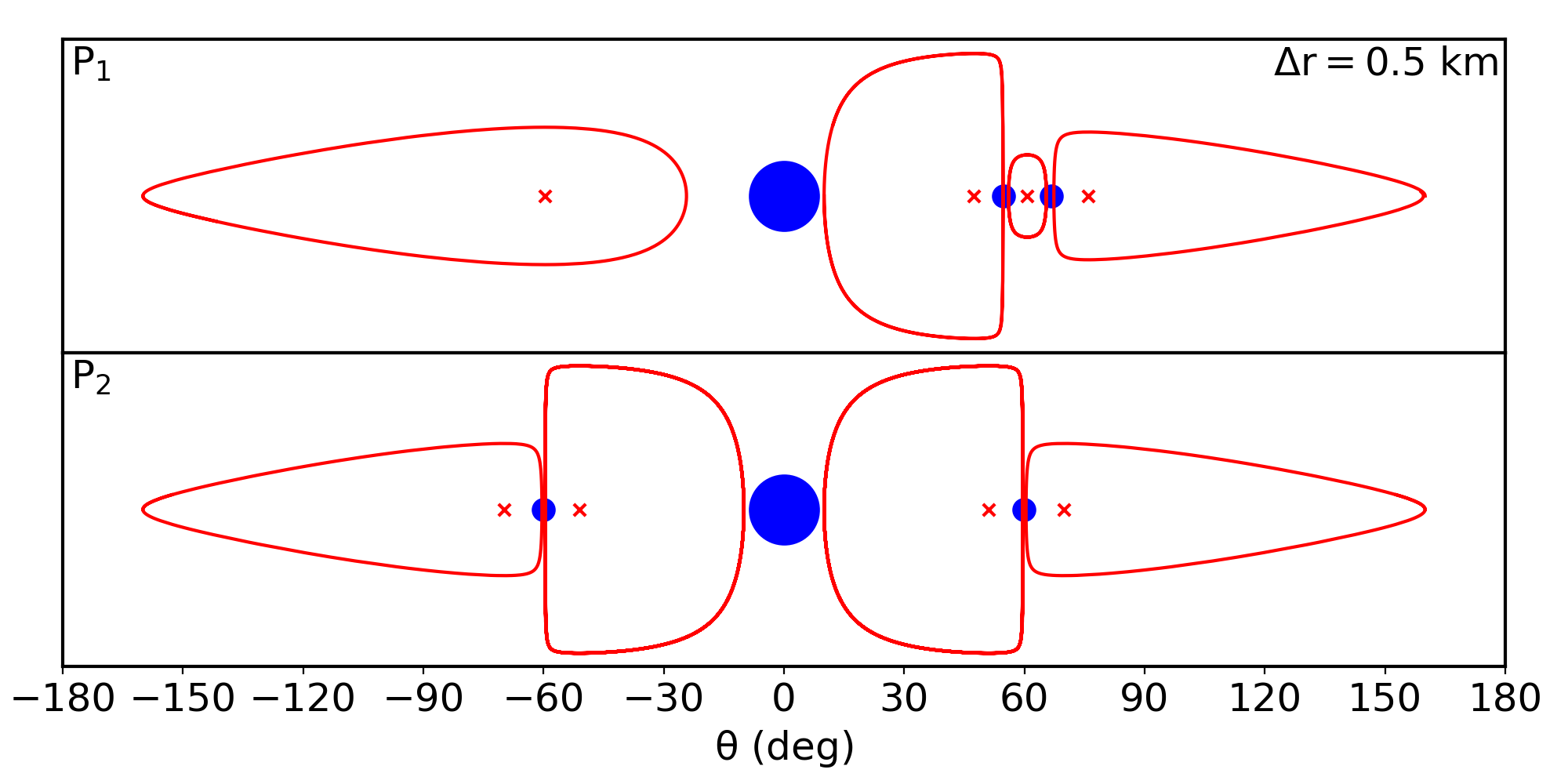}
\caption{Same as Figure~\ref{ptos2p} for 1+2 co-orbital satellite system.}
\label{ptos3p}
\end{figure}
\begin{figure}
\centering
\includegraphics[width=\columnwidth]{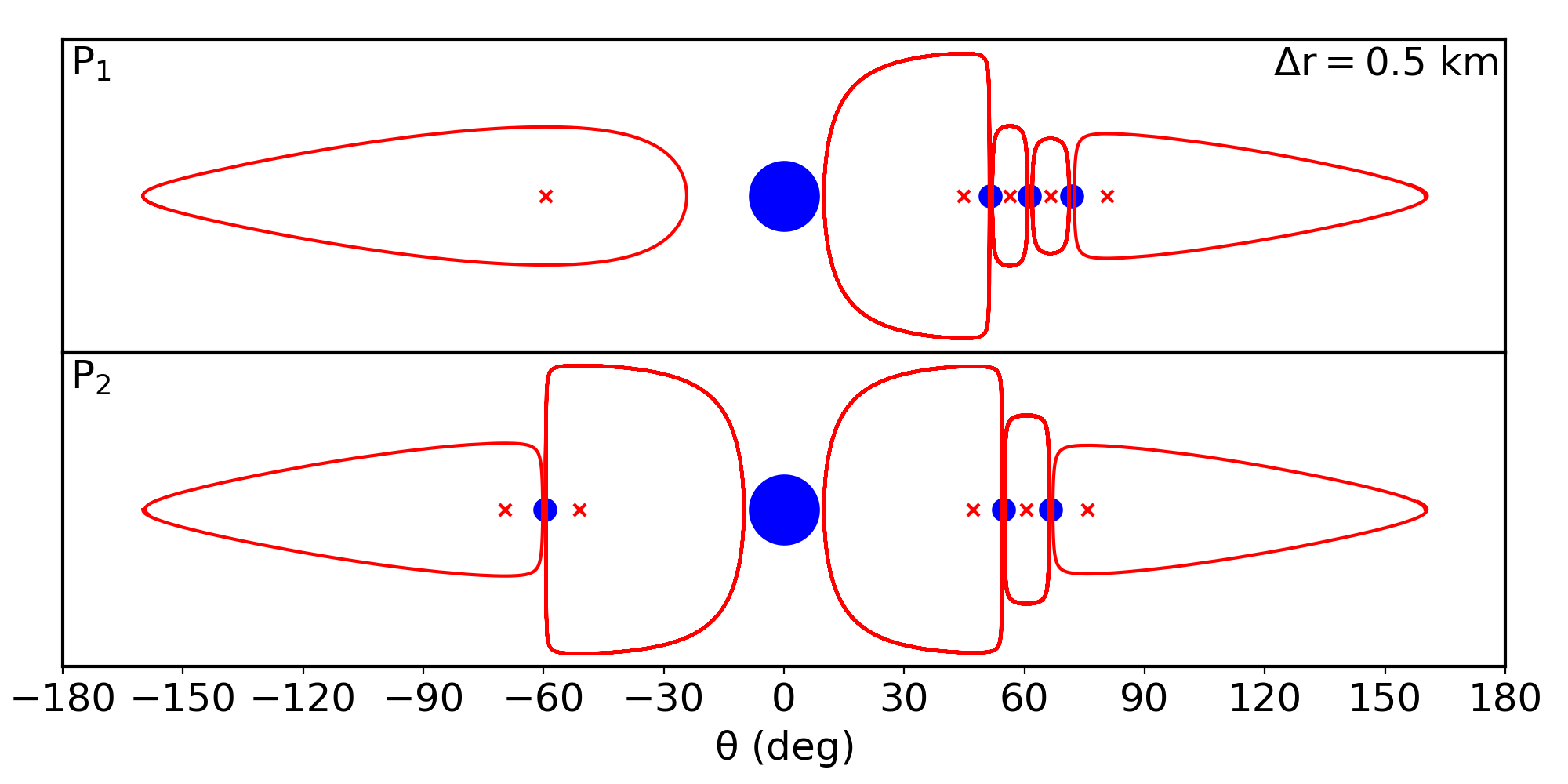}
\caption{Same as Figure~\ref{ptos2p} for 1+3 co-orbital satellite system.}
\label{ptos4p}
\end{figure}
\begin{figure}
\centering
\includegraphics[width=\columnwidth]{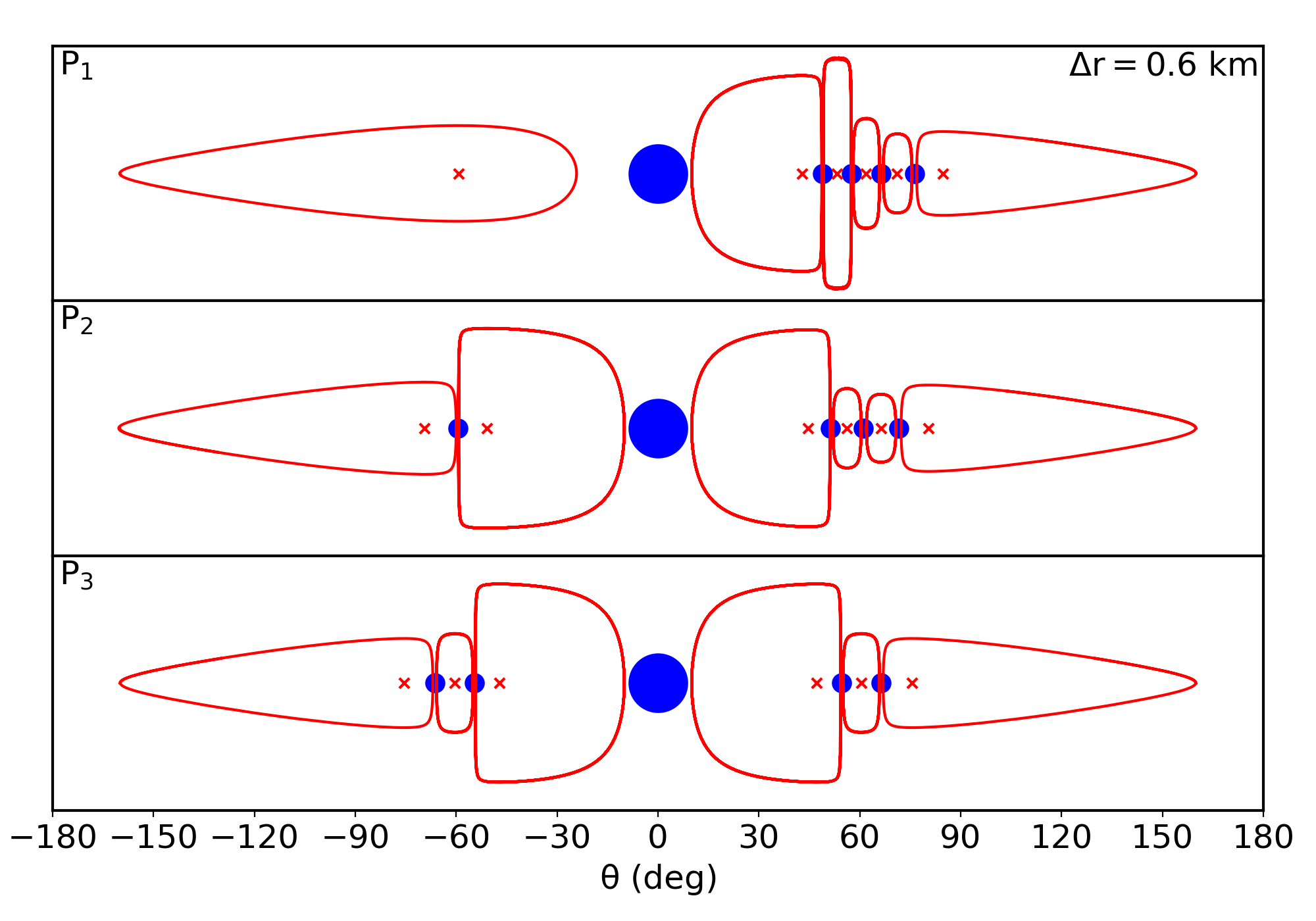}
\caption{Same as Figure~\ref{ptos2p} for 1+4 co-orbital satellite system.}
\label{ptos5p}
\end{figure}

The equilibrium configurations of the 1+$N$ co-orbital satellite system for $N$=1, 2, 3, and 4 are presented in Figures~\ref{ptos2p}, \ref{ptos3p}, \ref{ptos4p}, and \ref{ptos5p}, respectively. Each system with $ N $ odd has $(N+1)/2$ equilibrium configurations asymmetric with respect to $S_1$, while systems with $N$ even have $(N+2)/2$ equilibrium configurations, being one of them symmetric \citep{Re04}. The large blue dot at $\theta$=0~deg is the moon, and the smaller ones are the moonlets. The red crosses are the equilibrium positions of particles, obtained in Section~\ref{spp}, and the red lines are the trajectories of representative particles. The $y$-axis gives the radial variation of the trajectories. In the figures' upper right corner, we have the amplitude $\Delta r$ of the trajectory with the largest radial variation ($y$-axis scale). For visual purposes, the moon/moonlets are not in scale. Just to be clear, the particle trajectory never crosses the satellite.

The nomenclatures ``${\rm P_i}$'' on each line correspond to the label we'll use to refer to each equilibrium configuration from now on. We start in the configurations with all moonlets grouped near the Lagrangian point $L_4$ and ends in the configurations with moonlets on either side of $S_1$, near $L_4$ and $L_5$. We point out that the mirror version of every asymmetric configuration is an equally possible solution. An example is $L_5$, corresponding to the mirrored version of $L_4$. 

We obtain that the small eccentricity of the system does not significantly alter the equilibrium positions of the system in comparison to the circular case. The positions found by us are the same as those obtained by \cite {Re04}. If we assume larger eccentricities ($e\sim10^{-3}$), however, the locations of the equilibrium positions will change.

\subsection{Particle stable equilibrium positions} \label{spp}
We obtain the particle equilibrium positions in a 1+$N$ co-orbital satellite system using a practical method, seeking the azimuthal location with maximum radial variation of particles, in a simulation with a set of randomly distributed co-orbital particles without the non-conservative term. The particle equilibrium positions are the maximum of the potential exerted by the co-orbital satellites, and for a 1+$N$ co-orbital satellite system, there will be 2+$N$ particle equilibrium positions. As a rule, the particles are confined at one of the moon's triangular points or azimuthally trapped between a pair of moonlets.

\begin{figure}
\centering
\includegraphics[width=\columnwidth]{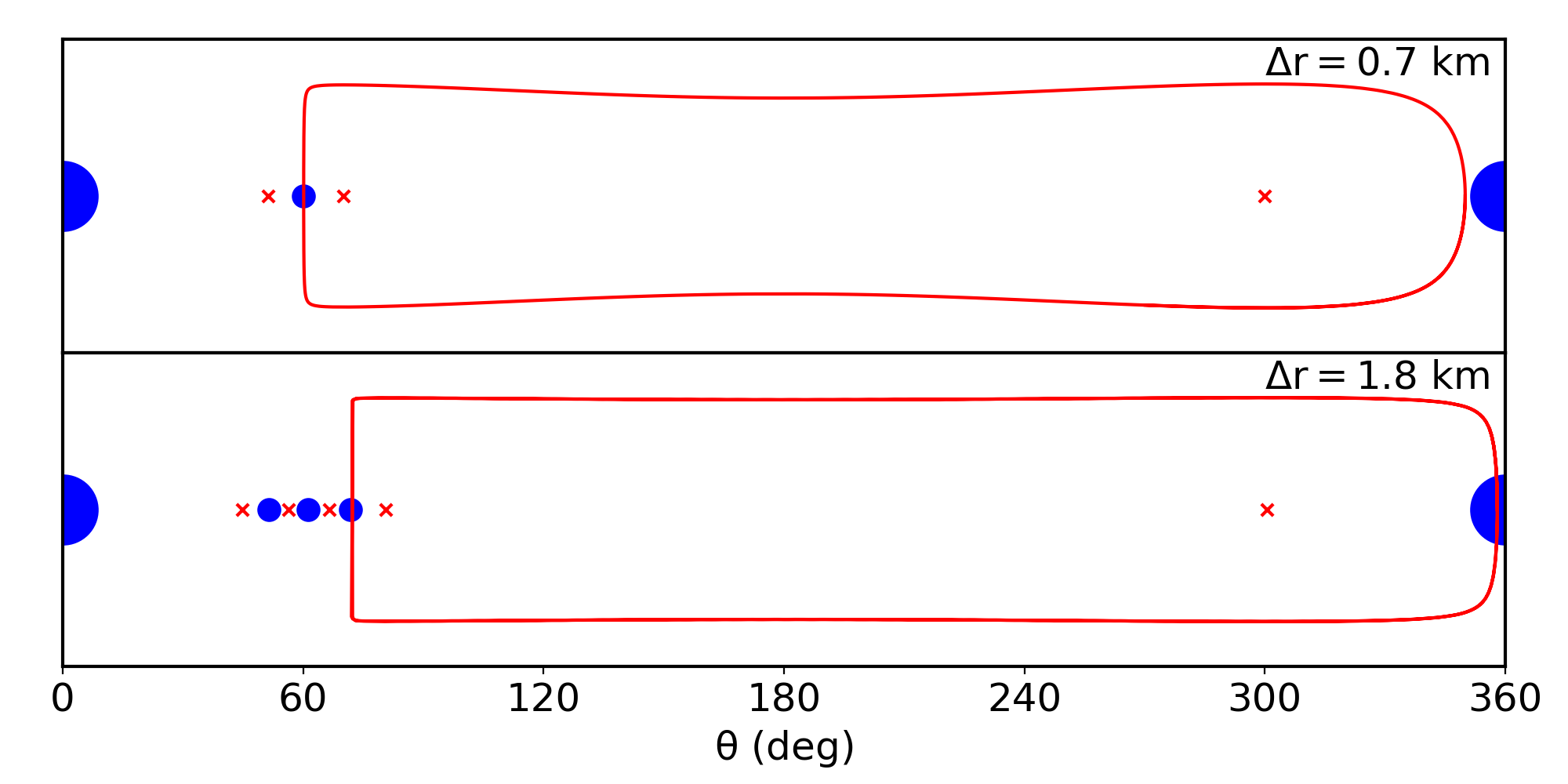}
\caption{Trajectory of a particle in horseshoe fashion orbit for a 1+1 (top panel) and 1+3 (bottom panel) co-orbital satellite system, both in $P_1$ configuration. The blue dots provide the location of moon/moonlets, and the red crosses are the particle equilibrium positions.}
\label{horseshoe}
\end{figure}

In Figures~\ref{ptos2p}-\ref{ptos5p}, the red crosses give the particle equilibrium positions, while the red lines are trajectories of some representative particles. All trajectories involve only one red cross and therefore correspond to tadpole-fashion orbits. This type of orbit is thought to be associated with the Neptune arcs, in the model of \cite{Re14}. There are also orbits involving more than one particle equilibrium position, as shown in Figure~\ref{horseshoe} for a 1+1 and 1+3 co-orbital satellites, from top to bottom. They correspond to horseshoe-fashion orbits and are beyond the scope of this work. In the next section, we give a step further by including the internal satellite Galatea into the system.

\section{Effects of Galatea on 1+$N$ co-orbital satellite systems} \label{lersys}

\begin{figure}
\subfigure[]{\includegraphics[width=\columnwidth]{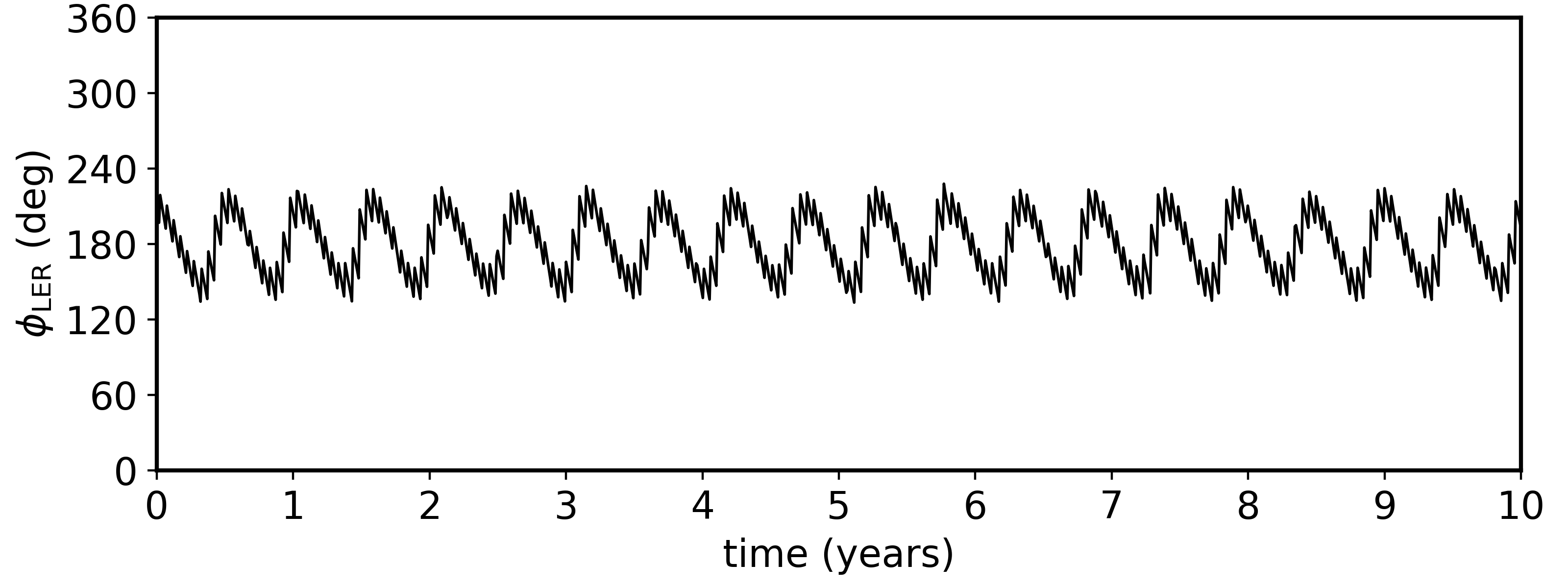}}
\subfigure[]{\includegraphics[width=\columnwidth]{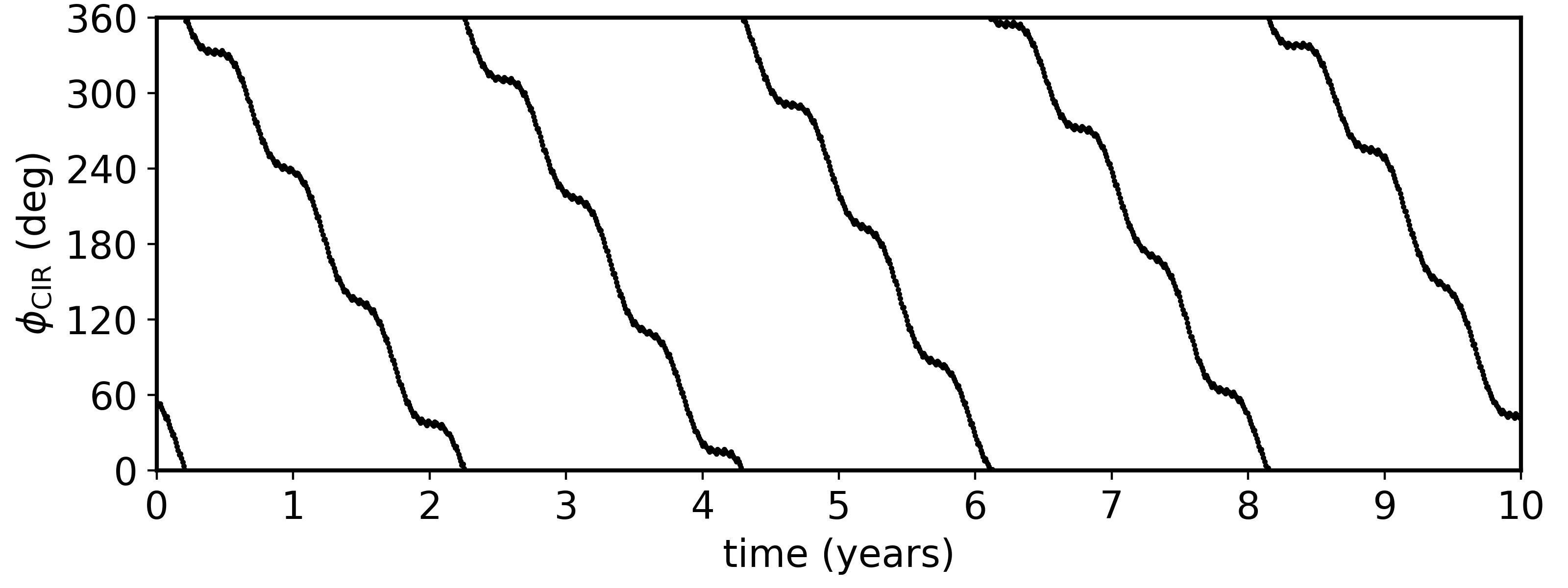}}
\caption{Resonant angles a) of 42:43 LER ($\phi_{\rm LER}$) and b) 84:86 CIR ($\phi_{\rm CIR}$) between $S_1$ and Galatea. The angles are given by $\phi_{\rm LER}=43\lambda-42\lambda_{\rm G}-\varpi$ and $\phi_{\rm CIR}=86\lambda-84\lambda_{\rm G}-2\Omega_{\rm G}$, where $\lambda$, $\varpi$, and $\Omega$ are mean longitude, longitude of pericentre, and argument of longitude node, respectively. The subscript $_{\rm G}$ refers to the satellite Galatea.}
\label{ler}
\end{figure}

Once we studied the equilibrium configurations in the co-orbital satellite system, we redid the simulations of Section~\ref{dyn} including the gravitational effects of Galatea with initial orbital elements given in Table~\ref{initial}. In this case, $S_1$ is involved in 42:43 LER with Galatea and displaced less than 1~km from 84:86 CIR \citep{Re14,Gi20}. The resonant angles associated with these resonances are shown in Figure~\ref{ler}. We used the algorithm presented in \cite{Re06} to transform the state vector into geometric orbital elements.

\begin{figure}
\centering
\includegraphics[width=\columnwidth]{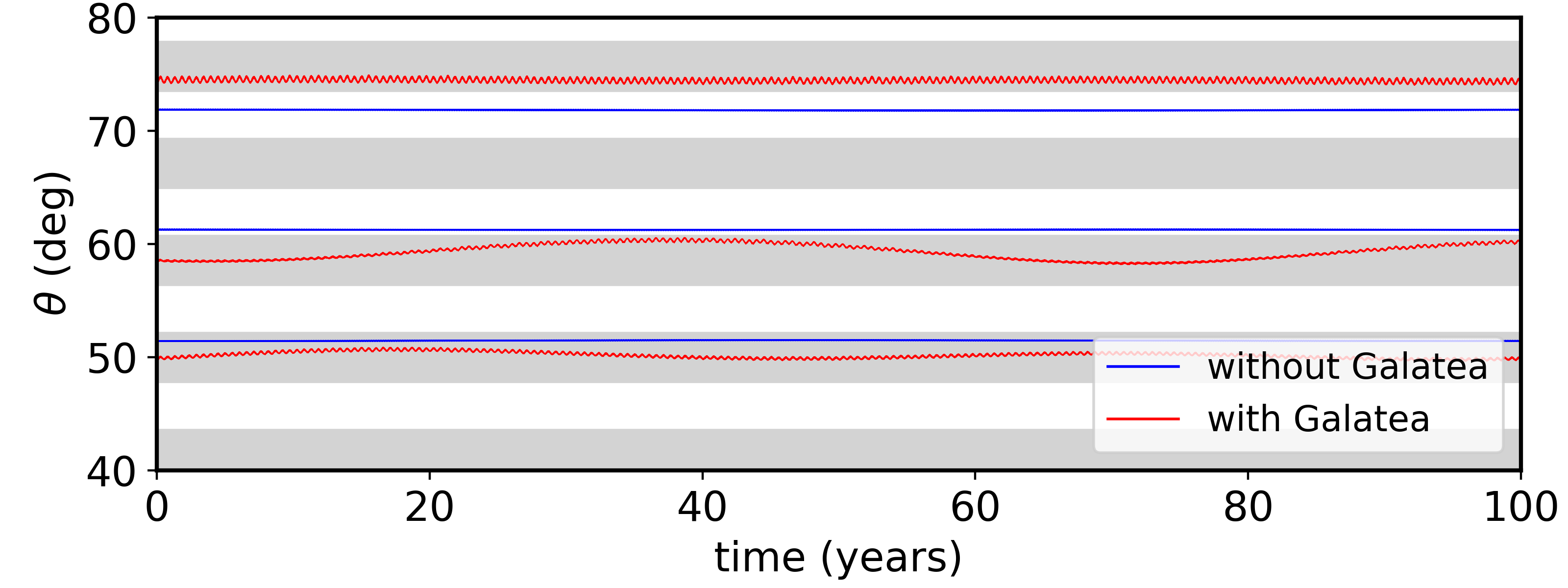}
\caption{Blue lines are the location of the equilibrium points for ${\rm P_1}$ configuration with 1+3 co-orbital satellites without Galatea's effects. Regions where one body remains in 42:43 LER with Galatea are shown as grey bands. The azimuthal evolution of the moonlets under the effects of Galatea is shown by the red lines.}
\label{e012}
\end{figure}

When Galatea is included in the system, we see a slight shift in the moonlet's equilibrium locations. Figure~\ref{e012} shows the longitudinal evolution of the moonlets in ${\rm P_1}$ configuration with 1+3 co-orbital satellites, for cases with and without Galatea (in red and blue, respectively). The grey bands (``LER bands'') correspond to the regions where the $\phi_{\rm LER}$ of a particle librates. As can be seen, the equilibrium location of the orbits shifts to the nearest LER band and the moonlets remain in a libration motion around the equilibrium. In this way, the moonlets remain confined azimuthally and radially around these ``new equilibrium positions''.

\begin{figure}
\centering
\includegraphics[width=\columnwidth]{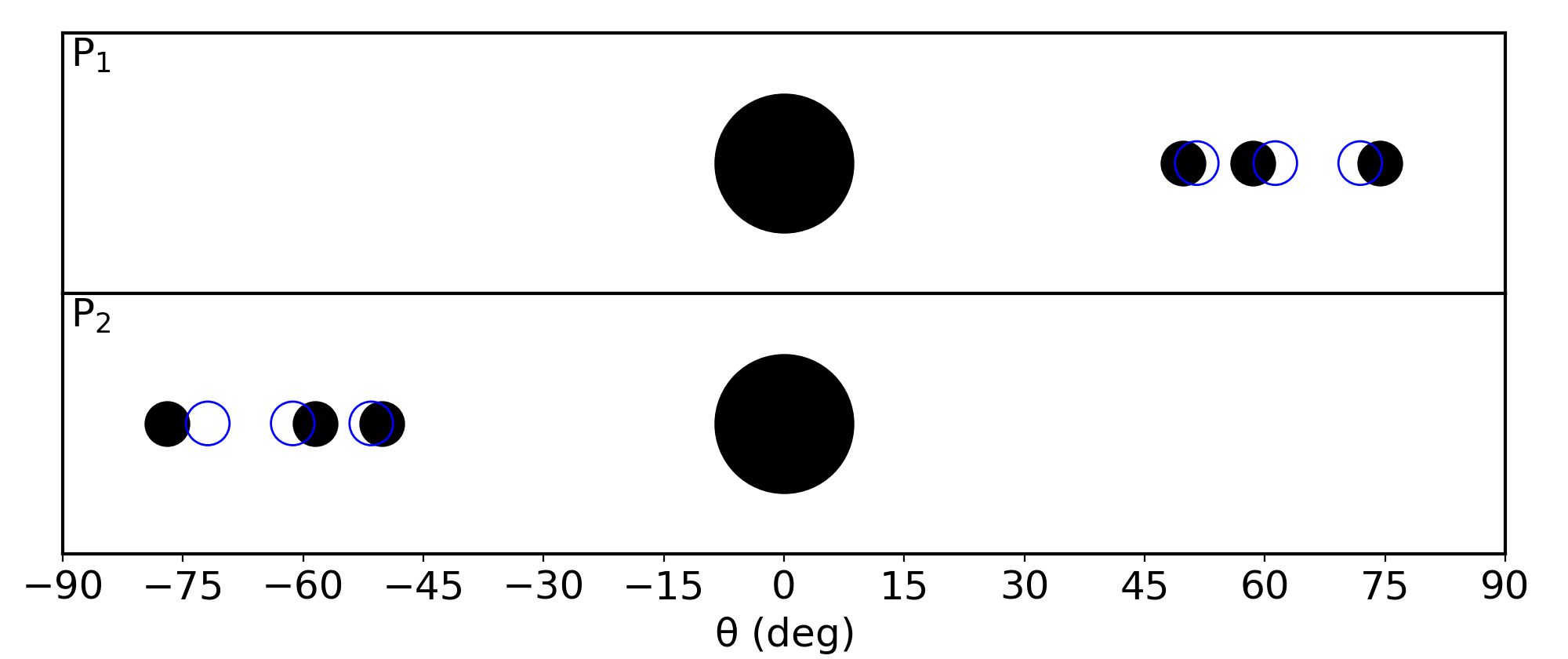}
\caption{Moonlet equilibrium positions in a 1+3 co-orbital satellite system. The black dots give the positions of the moonlets when we include Galatea. Unfilled blue dots correspond to the case without the satellite.}
\label{ptos4pcg}
\end{figure}

Figure~\ref{ptos4pcg} gives the moonlet equilibrium positions (black dots) for the 1+3 co-orbital satellite system, under Galatea effects. The equilibrium positions for the case without the satellite are the unfilled blue dots. If we assume different conditions for the moon/Galatea, the initial condition in the resonance phase space will change, leading to slightly different effects on moonlets motion \citep{Fo96}. As a consequence, the angular positions of the moonlets will differ by a few degrees. The displacement of the equilibrium locations due to LER were also verified by \cite{Fo96} and \cite{Mo13} in systems where the overlap of corotation and Lindblad resonances comes from the same satellite. In our work, a similar phenomenon occurs, with the difference that corotation and Lindblad resonances originate from different bodies.

\begin{figure}
\centering
\subfigure[]{\includegraphics[width=\columnwidth]{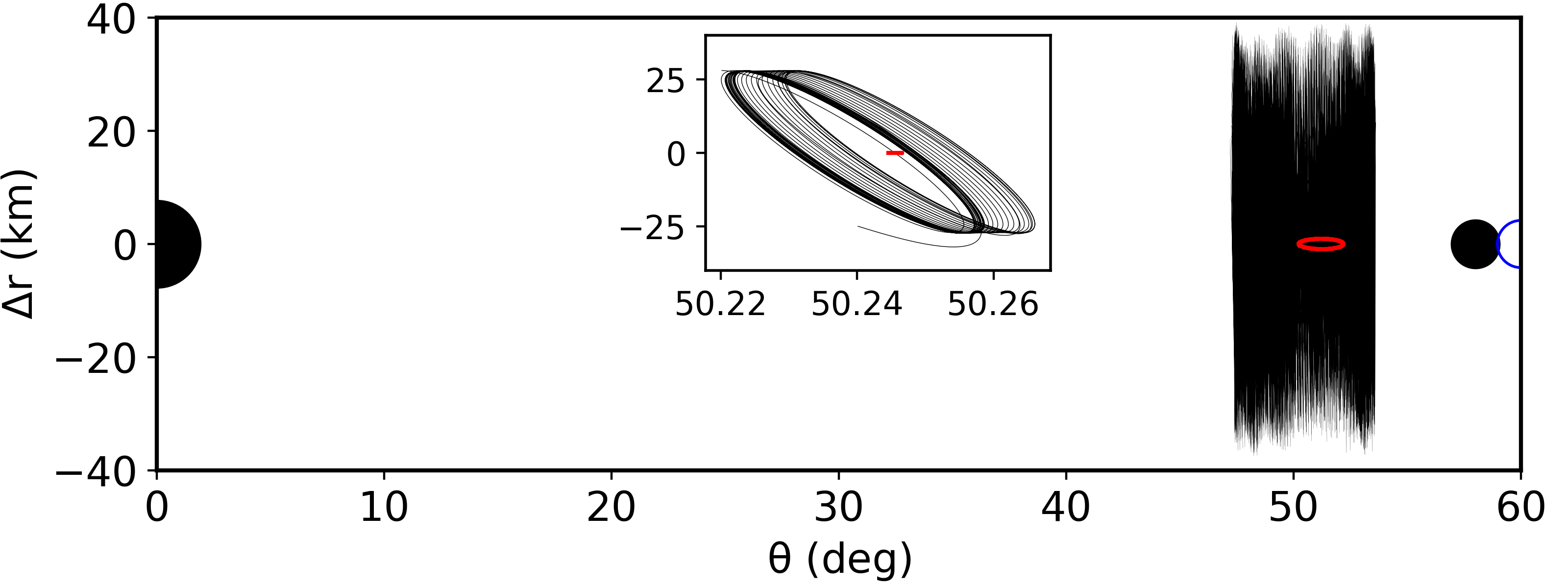}}
\subfigure[]{\includegraphics[width=\columnwidth]{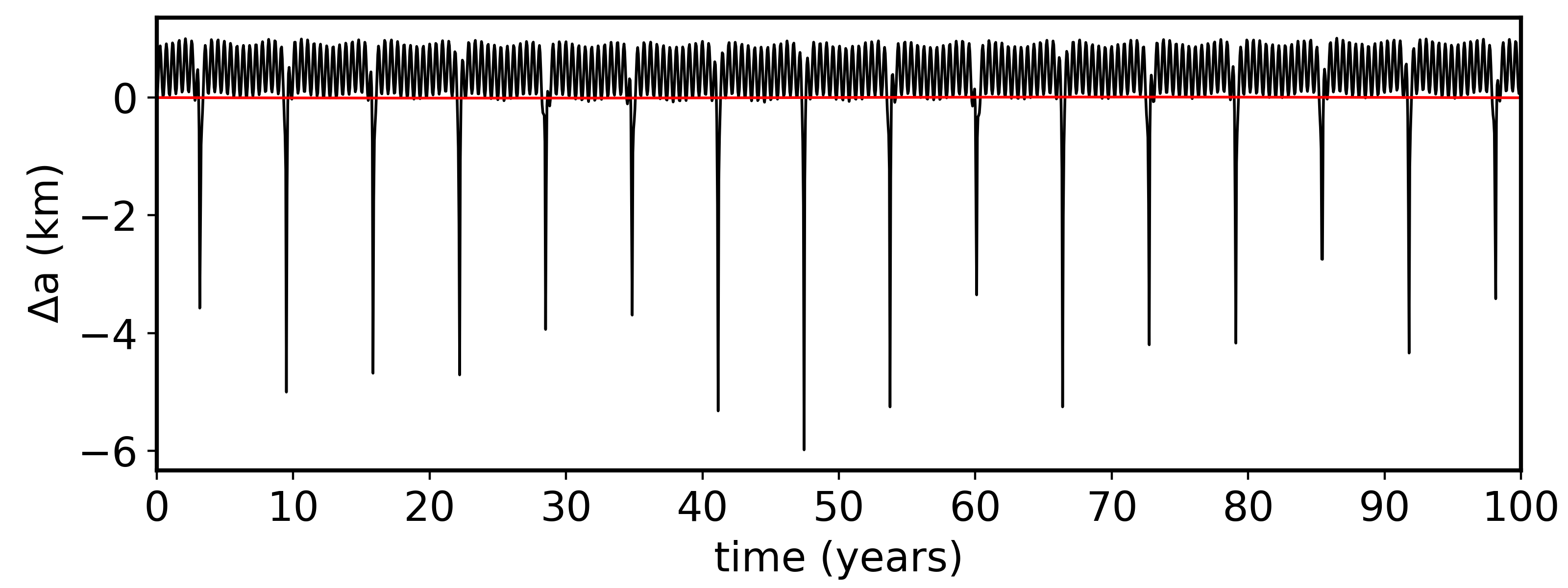}}
\subfigure[]{\includegraphics[width=\columnwidth]{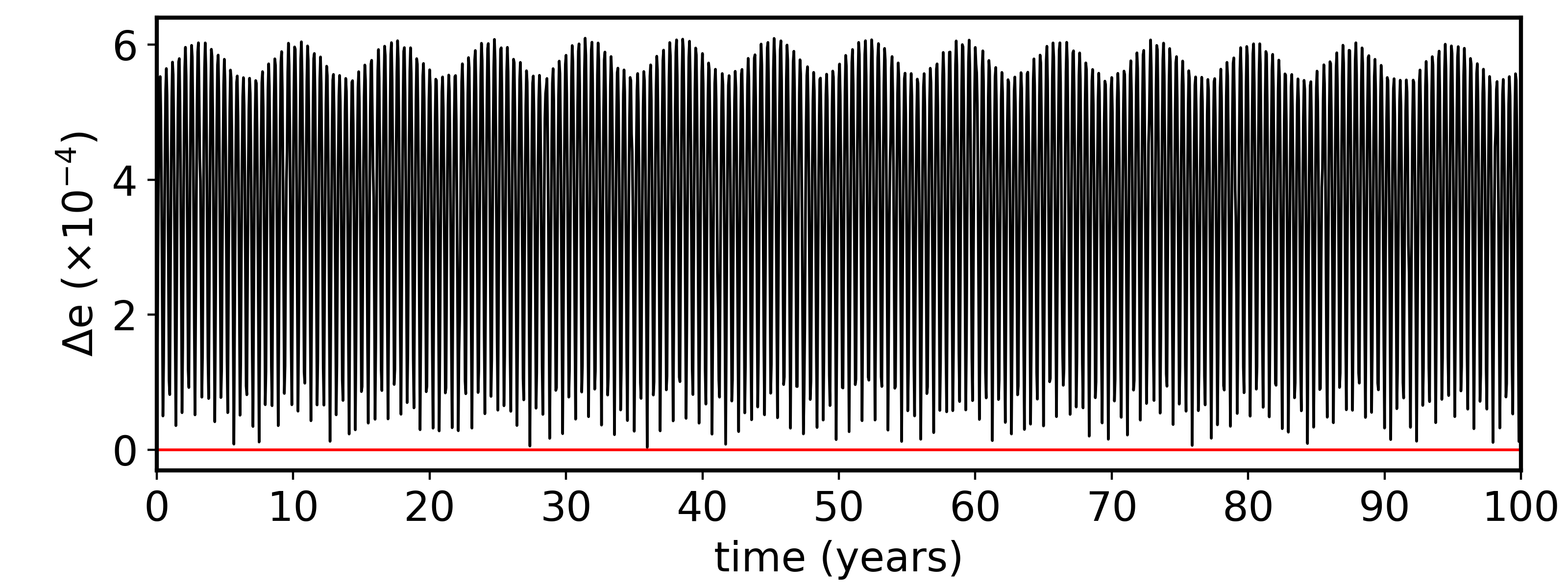}}
\caption{a) Azimuthal and radial variation, b) semi-major axis, and c) eccentricity of a representative particle in a ${\rm P_1}$ configuration with 1+1 co-orbital satellites. Solid black line provides the particle in the system under the effects of Galatea, and the red line is the case without the satellite. The simulation timespan is 100~years, with the first 50~days shown in the zoom. In the top panel, the moonlet confining the particle is in black and blue for the case with and without Galatea, respectively.}
\label{pcsg}
\end{figure}

Galatea also disturbs the orbital evolution of the particles, which oscillate around the equilibrium positions with larger radial variations. Figure~\ref{pcsg} shows the radial variation (top panel), semi-major axis (middle panel), and eccentricity (bottom panel) of a representative particle in ${\rm P_1}$ configuration with 1+1 co-orbital satellites, for 100~years. The timespan of the zoom box in the top panel is of 50~days. Cases with and without Galatea are in black and red lines, respectively.

In the top panel, the black dot places the moonlet equilibrium position in the system with Galatea, and the unfilled blue dot places the position in the system without the satellite. As can be seen in the zoom, the particle shows an additional oscillation with Galatea in the system, which translates into larger radial and azimuthal variations in the particle motion. The satellite is also responsible for the larger variations in the semi-major axis and eccentricity of the particle. The peaks seen in the temporal evolution of the semimajor axis are due to closest approach between the particle and the co-orbital satellite.

\begin{figure}
\centering
\subfigure[]{\includegraphics[width=\columnwidth]{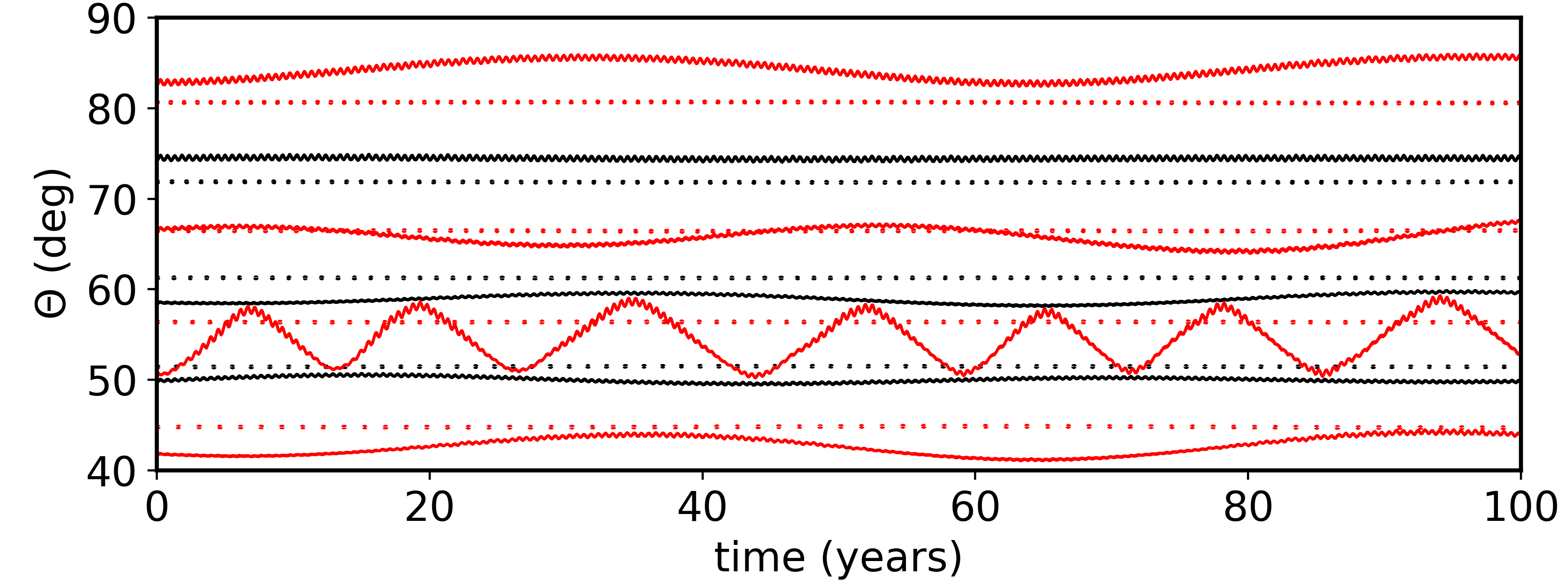}}
\subfigure[]{\includegraphics[width=\columnwidth]{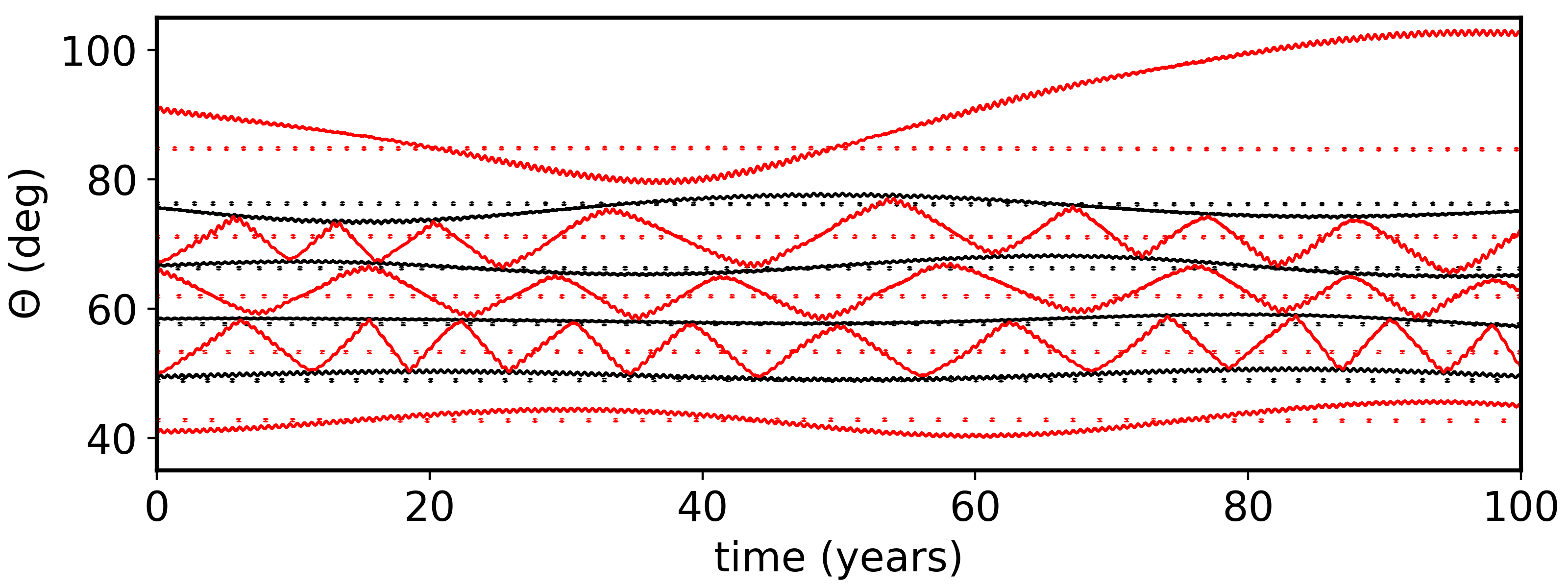}}
\subfigure[]{\includegraphics[width=\columnwidth]{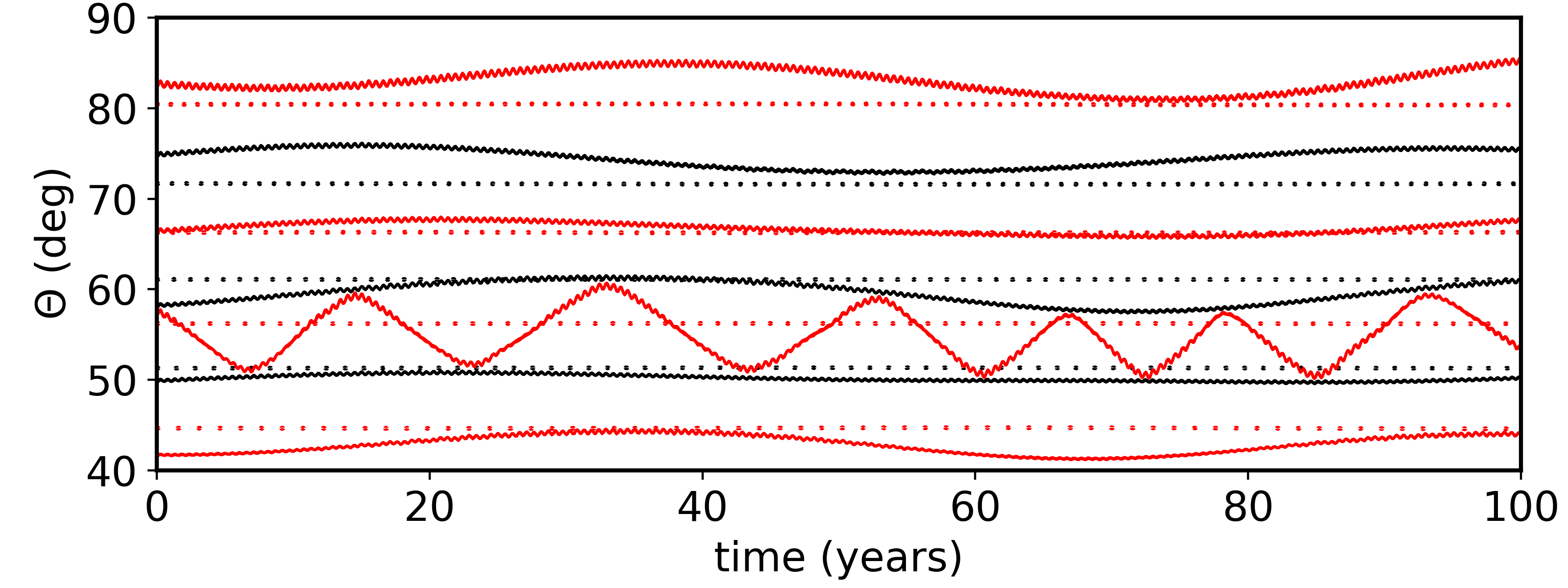}}
\caption{Temporal variation of the azimuthal angle of moonlets (black lines) and particles (red lines) for a) ${\rm P_1}$ configuration with 1+3 co-orbital satellites, b) ${\rm P_1}$ configuration with 1+4 co-orbital satellites, and c) ${\rm P_2}$ configuration with 1+4 co-orbital satellites. The full lines correspond to the case with Galatea, and the dotted lines are the trajectories for the case without the satellite.}
\label{partsL4}
\end{figure}

To exemplify equilibrium configurations when Galatea is presented in the system, we show in the Figure~\ref{partsL4}, from top to bottom, the ${\rm P_1}$ configuration with 1+3 co-orbital satellites and the ${\rm P_1}$ and ${\rm P_2}$ configurations with 1+4 co-orbital satellites. These cases are especially interesting because they can reproduce the angular distribution of the arcs Fraternit\'e, Egalit\'e, Libert\'e, and Courage. In the figure, the particles (in red) are azimuthally confined by the moonlets (in black). Solid lines correspond to the case with Galatea, and the dotted lines to the case without the satellite. In the next section, we simulate a set of fragments supposedly formed in the disruption of an old moon, analysing whether they can give rise to a system of 1+N co-orbital satellites.

\section{Temporal evolution of fragments from a moon disruption} \label{4frag}

\subsection{Impact between an ongoing object and a trojan moon}

\begin{figure*}
\subfigure[Collisional event]{\includegraphics[width=0.66\columnwidth]{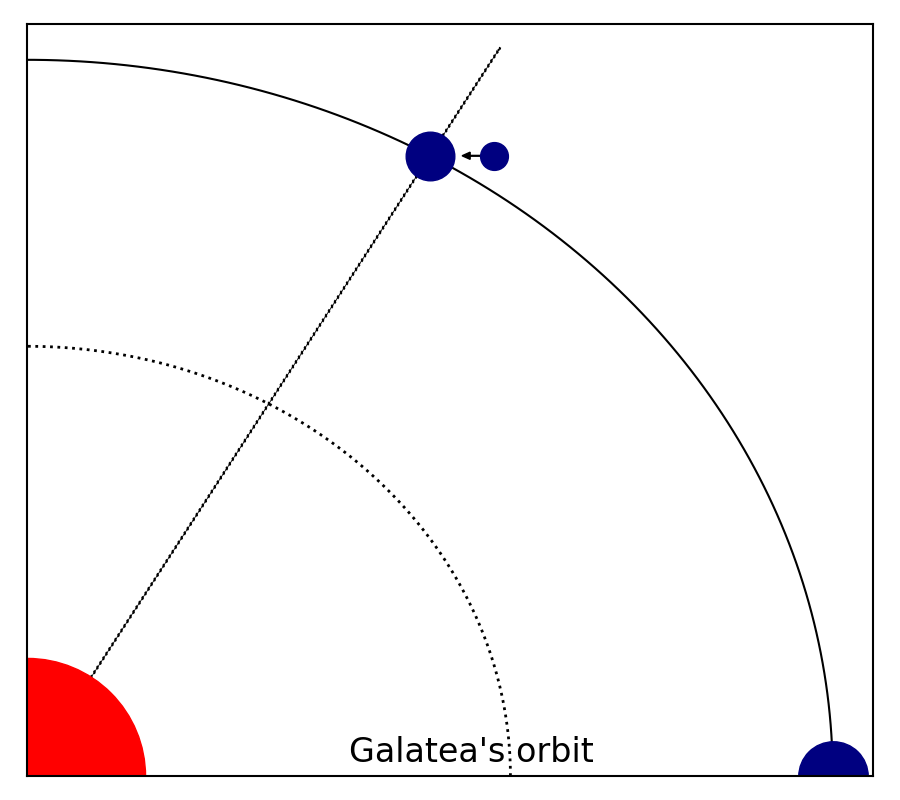} \label{pa}}
\subfigure[Disruption into fragments]{\includegraphics[width=0.66\columnwidth]{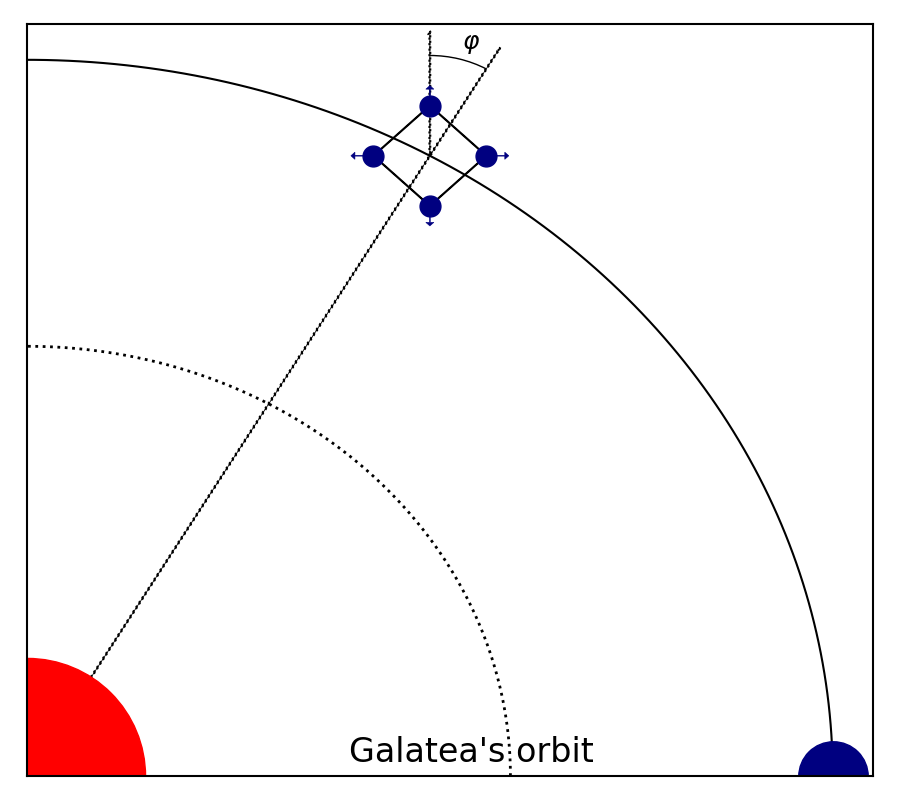} \label{pb}}
\subfigure[Moonlets in an equilibrium configuration]{\includegraphics[width=0.66\columnwidth]{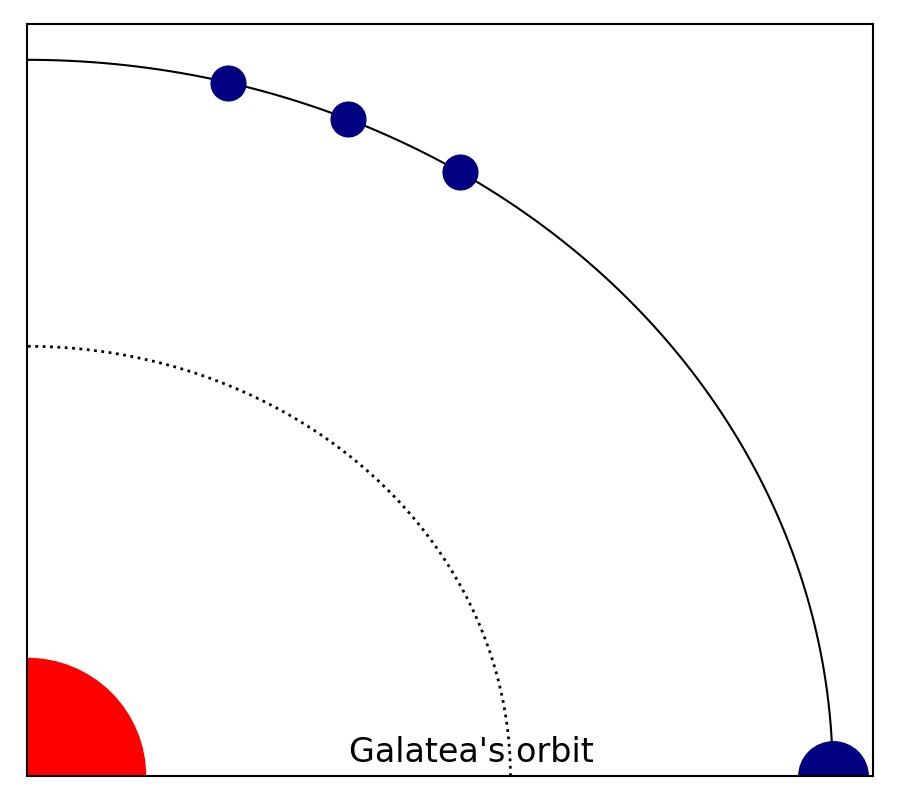} \label{pc}}
\caption{Steps from the formation of co-orbital moonlets: After a collisional event, an ancient body located at the Lagrangian point of $S_1$ disrupts into a set of fragments. The fragments collide and form moonlets that settle in the equilibrium positions.}
\label{tsteps}
\end{figure*}

In light of the Janus/Epimetheus formation model proposed in \cite{Tr15}, we envision the following scenario for the formation of a 1+$N$ co-orbital satellite system (Figure~\ref{tsteps}): Initially, we assume an ancient system composed of the moon $S_1$ ($R_{\rm S_1}=5.2$~km) and an object located at one of its triangular points (trojan). After an impact with an ongoing object (Figure~\ref{tsteps}a), the trojan disrupts, forming a set of fragments (Figure~\ref{tsteps}b). The fragments perform horseshoe orbits with $S_1$ and collide with each other, giving rise to moonlets. Finally, the moonlets settle into equilibrium positions of the system (Figure~\ref{tsteps}c).

Keeping in mind the system proposed by \cite{Re14}, we define the minimum trojan mass as $m_{\rm tro}=4\times 10^{-2}m_{S_1}$, corresponding to an object made of ice with physical radius of $R_{\rm tro}=1.8$~km. $m_{\rm tro}$ is approximately the sum of the masses of the moonlets $S_2$, $S_3$ and $S_4$ proposed by \cite{Re14}. For a fiducial impact of ${\rm 3000~m/s}$, the minimum incident kinetic energy per mass $Q_*$ required to disrupt the trojan is \citep{Be99}
\begin{equation}
Q_*=2.7\times 10^{-12}\left(\frac{R_{\rm tro}}{1~{\rm m}}\right)^{-0.39}+4\times10^{-2}\left(\frac{R_{\rm tro}}{1~{\rm m}}\right)^{1.26}~{\rm J/kg}.
\end{equation}
The radius $R_{\rm imp}$ required for an ice impactor to disrupt the trojan can be estimated as \citep{St12,Me17}
\begin{equation}
R_{\rm imp}=\left(\frac{3}{2\pi(10^3~{\rm kg/m^3})}\frac{Qm_{\rm tro}}{(3000~{\rm m/s}+v_{\rm esc})^2}\right)^{1/3}   
\end{equation}
where $v_{\rm esc}$ is the escape velocity of the trojan and $Q$ is the reduced kinetic energy of the system. For a disruption, $Q\geq Q_*$.

\cite{Be99} shows that the mass of the largest remnant $m_{\rm lr}$ produced by the disruption of the trojan can be estimated as
\begin{equation}
m_{\rm lr}=0.5-0.6\left(\frac{Q}{Q_*}-1\right)m_{\rm tro}.
\end{equation}
From these relations we find, for example, that a kinetic energy $Q/Q_*=1.4$ and an impactor of $R_{\rm imp}\approx100$~m are needed for the trojan to be destroyed, and its largest remnant has the mass $m_{\rm lr}=m_{\rm tro}/4$.

Neptune's sphere of influence is regularly crossed by comets from the Kuiper belt. It was proposed by \cite{Co92} that such objects were responsible for catastrophic disruptions of original moons. These events would be the sources of ancient rings around the planet. \cite{Le00} calculated the comet impact rate in the Neptune region as $3.5\times 10^{-4}$yr$^{-1}$. Taking into account the gravitational focusing \citep{Le97,St00}, we obtain that a comet reaches the Adams ring region every $\sim10^5$~yrs. Since these comets have typical sizes \citep[${\rm \sim km}$,][]{Le00} larger than $R_{\rm imp}$, it seems that a trojan disruption, caused by impacts with such comets, is possible.

If we assume that the arcs are composed of particles with physical radius $s$ ranging from ${\rm 1~\mu m}$ to ${\rm 1~m}$, following a numerical distribution given by $dN\propto s^{-3.5}ds$ \citep{Co92}, we get a total mass of $\sim 4\times 10^{10}$~kg for the structures \citep{sfair2012role,Gi20}. This value is an order of magnitude greater than the mass of the ongoing object, which means that a $100$~m-sized impactor does not contain the material needed to fill the observed arcs, requiring additional material production mechanisms. We will discuss some of these processes in the next section.

A disruption is an extreme event, responsible for producing numerous fragments. However, the rupture of an object of a few kilometres due to an impact of ${\rm \sim~km/s}$ usually gives rise to a limited number of larger fragments with the same mass order (kilometric fragments), while producing a large amount of material with sizes ranging from micrometres to metres \citep{Mi04,St09,Ju17}. In this section, we stick to calculate the evolution of these larger fragments, while some comments on the smaller fragments are addressed in Section~\ref{arcsloc}.

\subsection{Simulations of a representative case} \label{representative}

To assess whether the trojan disruption generates a family of co-orbital satellites, we performed a set of simplistic numerical simulations starting right after the trojan disruption. For this, we use the \textit{Mercury} package \citep{Ch99}, with the Bulirsch–Stoer algorithm. The dynamical system is composed of Neptune and its gravitational coefficients ($J_2$ and $J_4$), Galatea, the moon $S_1$, and the major fragments of the disruption. We also include the non-conservative term for carrying the fragments to the equilibrium positions.

Next, we present the results obtained in 3000 numerical simulations for a representative case with $\nu=10^{-4}$~yr$^{-1}$ and four fragments of same mass ${\rm m_{\rm fra}=10^{-2}m_{S_1}}$ ($R_{\rm fra}\approx1.1$~km). The fragments are distributed at the vertices of a regular polygon with four sides centred on the $L_4$ point. The length of the polygon sides is $l=2R_{\rm fra}+100$~m, and its angular orientation with the radial direction ($\varphi$) is given randomly in the range $0$~deg-$180$~deg (Figure~\ref{tsteps}b).

From Figure~15 of \cite{Be99}, we obtain that fragments with mass $m_{\rm tro}/4$ are ejected with radial velocities $\sim 0.4$~m/s. With that in mind, we get randomly chosen ejection velocities from $0.36$~m/s to $0.73$~m/s. For velocities below $0.36$~m/s, the relative velocity between the fragments is very low, and they collide right at the beginning of the simulation. For velocities larger than $0.73$~m/s, the fragments leave the horseshoe region. 

\begin{figure}
\centering
\includegraphics[width=\columnwidth]{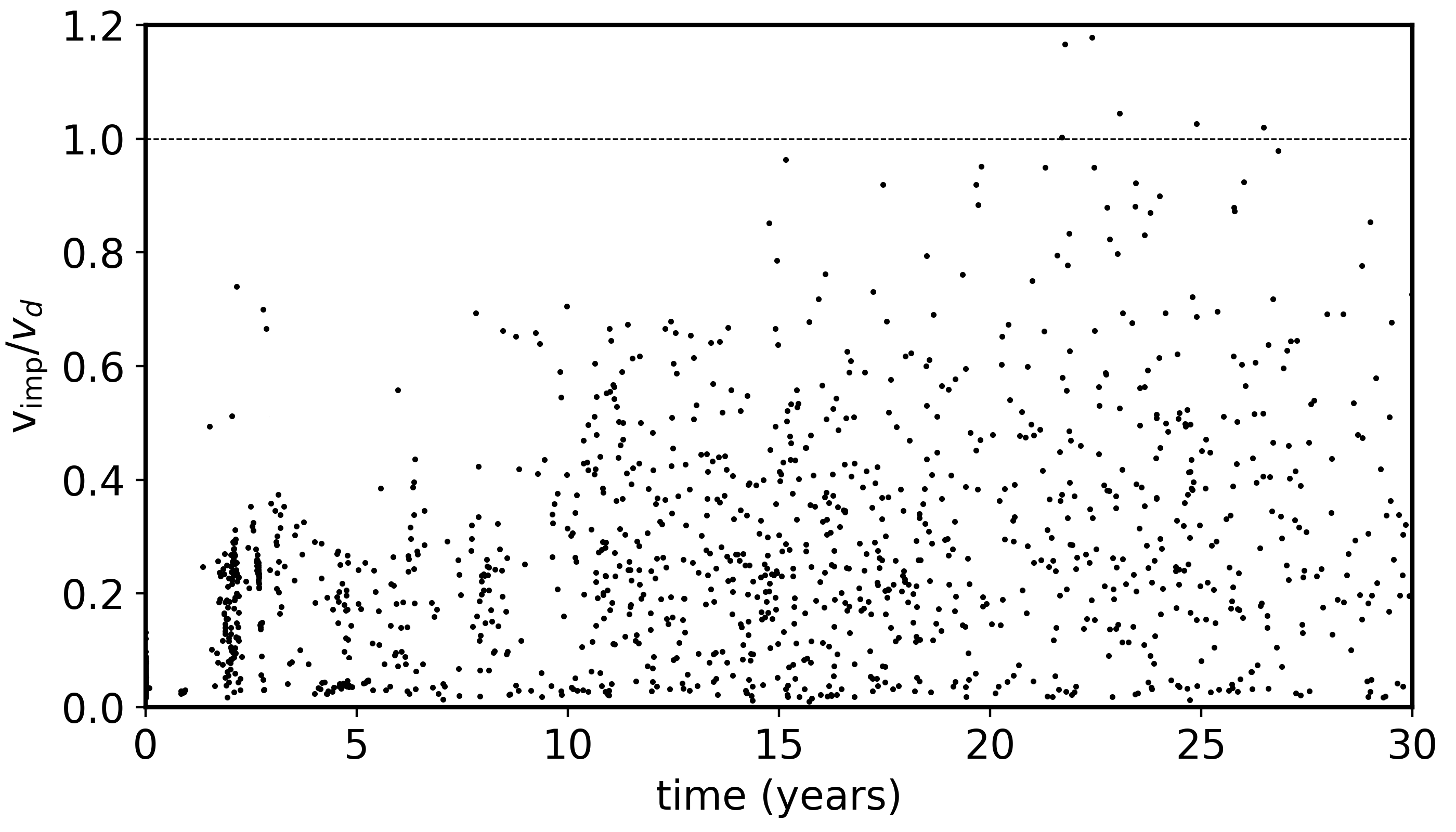}
\caption{Fraction between impact velocity and cut-off velocity as a function of the time the impacts occurred. The dotted vertical line provides the boundary between constructive and disruptive collisions.}
\label{collfrag}
\end{figure}

The \textit{Mercury} package treats collisions as inelastic events with conservation of linear momentum and mass. To ensure the validity of this treatment, we compare the impact velocity $v_{\rm imp}$ between the fragments with a cut-off velocity $v_d$ for which a collision can be considered as constructive, given by \citep{St12,Tr15}
\begin{equation}
v_d=\sqrt{2Q_*\mu_{\rm fra}}
\end{equation}
where the values of $Q_*$ are taken from Fig.~11 of \cite{St12} and $\mu_{\rm fra}$ is the mass ratio between the fragments ($\mu_{\rm fra}\geq1$). Figure~\ref{collfrag} shows $v_{\rm imp}/v_{d}$ as a function of time for our 3000 numerical simulations. As can be seen, only a small part of impacts ($<1\%$) have velocities above the cut-off limit, showing that collisions can be, in general, treated as constructive events

Despite this, the impacts have relatively high velocities, $v_{\rm imp}\sim$m/s (an order of magnitude higher than the ejection velocity). This fact is a result of the forced eccentricity gradient caused by the resonance with Galatea, as discussed in works such as \cite{Po95}, \cite{Fo96} and \cite{Re14}. The high impact velocities, especially those in the last years of simulation, indicate that the impacts are not always perfect merging, but events with partial merging or erosion, presenting themselves as possible sources for the arcs. Some comments about it are addressed in Section~\ref{arcsloc}.

\begin{figure}
\centering
\subfigure[]{\includegraphics[width=\columnwidth]{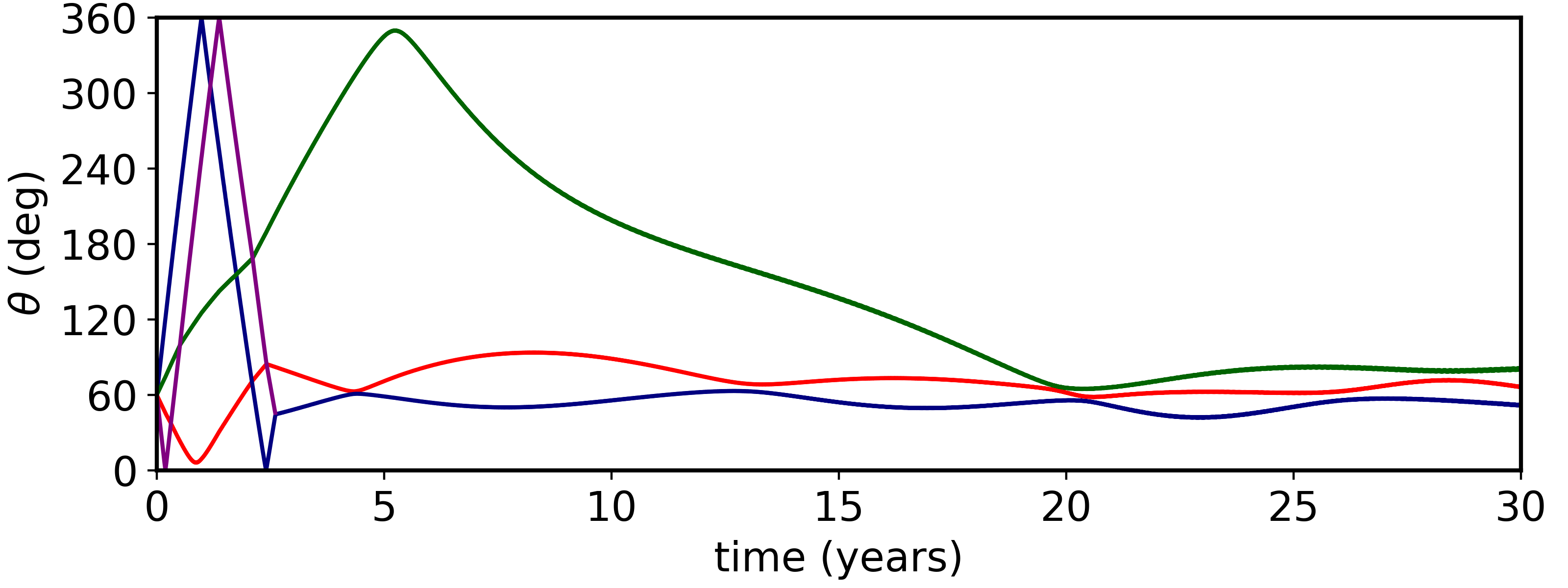} \label{figform3b}}
\subfigure[]{\includegraphics[width=\columnwidth]{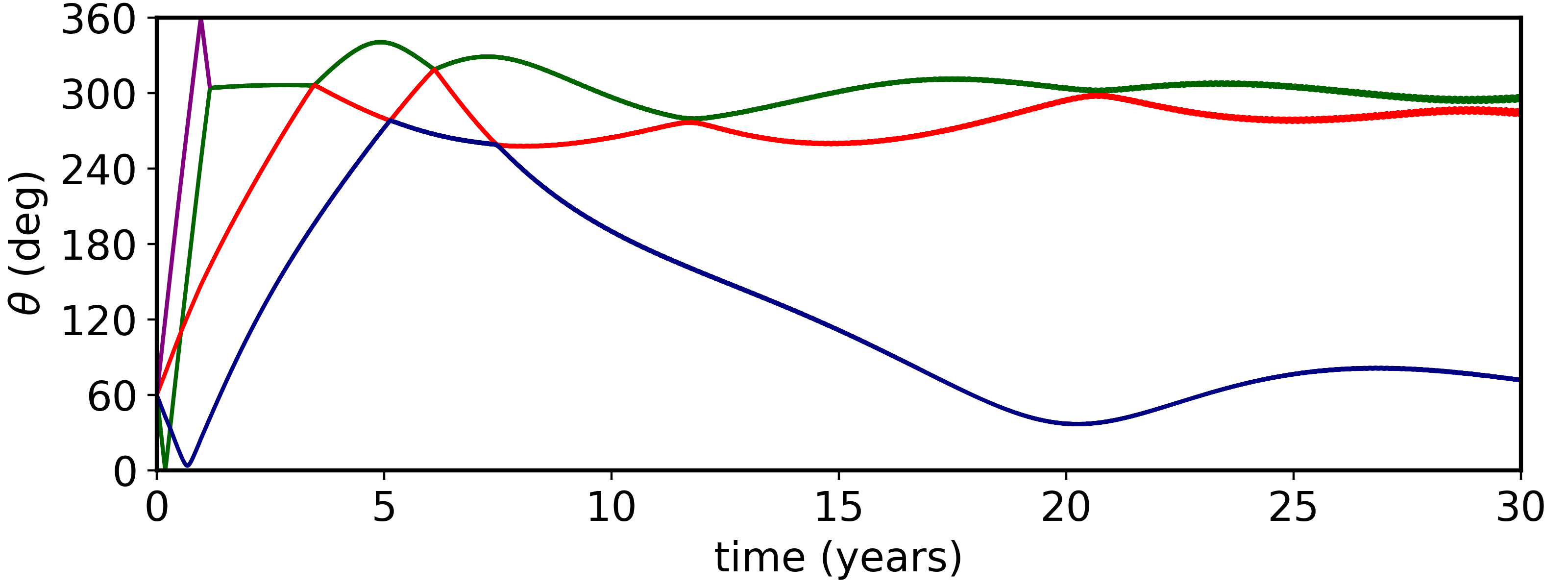} \label{figform3a}}
\caption{Angular evolution of fragments in two systems that form 1+3 co-orbital satellites. In both systems, two fragments collide, giving rise to a moonlet with mass ${\rm 2m_{\rm fra}}$ and a pair of moonlets with masses ${\rm m_{\rm fra}}$. We got a ${\rm P_1}$ final configuration in the top panel and a ${\rm P_2}$ configuration in the bottom. The ${\rm 2m_{\rm fra}}$ mass moonlet is shown in blue line in the top panel and in green line in the bottom one.}
\label{figform3}
\end{figure}

\begin{figure}
\centering
\includegraphics[width=\columnwidth]{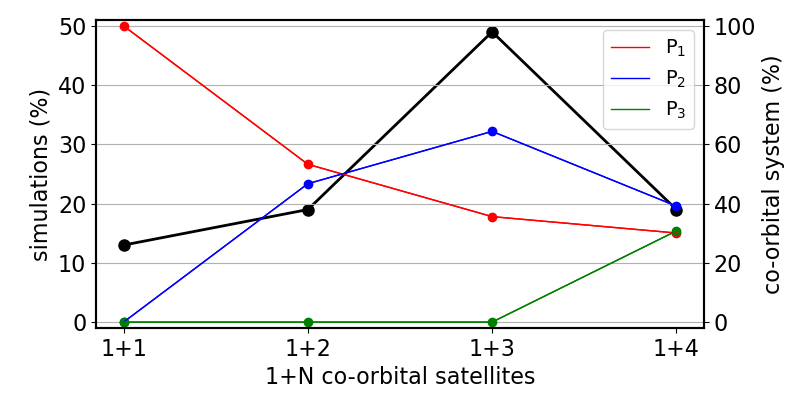}
\caption{Left y-axis gives the fraction of 1+$N$ co-orbital satellites obtained in the 3,000 numerical simulations (black line) while the fractions of systems in ${\rm P_i}$ configuration relative to each set of 1+$N$ co-orbital satellite system ($N=$1, 2, 3, and 4) are given on the right y-axis (coloured lines). The dynamical system includes four fragments of same mass (${\rm m_{\rm fra}=10^{-2}m_{S_1}}$), Neptune and its gravitational coefficients, Galatea, $S_1$, and a non-conservative term $\nu=10^{-4}$~yr$^{-1}$. \label{esrate}}
\end{figure}

We obtained that about 13\% and 19\% of the systems give rise to 1+1 and 1+2 co-orbital satellites, respectively (Figure~\ref{esrate}). About 49\% of the simulations form 1+3 co-orbital satellites, while in $\sim 19$\% of the simulations the fragments do not collide and form 1+4 satellites. We classify the formed systems in relation to their final equilibrium configurations through the classifications given in Figures~\ref{ptos2p}-\ref{ptos5p}. The 42:43 LER angle librates for all moonlets at the end of the simulations. Figure~\ref{figform3} shows examples of the formation of 1+3 co-orbital satellite systems, where in the top panel is a ${\rm P_1}$ configuration while in the bottom is presented a ${\rm P_2}$ configuration. 

Despite having the same ${\rm P_i}$ configuration, two systems can be dynamically different, depending on the moonlets mass distribution. For example, a ${\rm P_1}$ configuration with 1+3 co-orbital satellites is obtained if a fragment collides with $S_1$ -- forming three moonlets with masses ${\rm m_{\rm fra}}$ -- or if two fragments collide with each other -- forming two moonlets with masses ${\rm m_{\rm fra}}$ and one with ${\rm 2m_{\rm fra}}$. The latter can correspond to three different dynamical systems depending on the position of the moonlet with mass ${\rm 2m_{\rm fra}}$, resulting in a total of four degeneracies for the same ${\rm P_1}$ configuration with 1+3 co-orbital satellites. However, our results showed that these different mass distributions are only responsible for producing small differences in the orbital evolution of the moonlets. Equilibrium positions are approximately the same for all the degenerate cases.

The fractions of numerical simulations that produce 1+$N$ co-orbital satellite systems are given in Figure~\ref{esrate} by the black line, whose values are given on the left y-axis. The figure also shows by coloured lines the fractions of systems in ${\rm P_i}$ configuration. The values of the coloured lines are given on the right y-axis and are relative to the number of simulations with 1+N co-orbital satellites, for $N=$1, 2, 3, and 4. For example, 19\% of the simulations produce systems with 1+2 co-orbital satellites (570 simulations), $\sim54$\% of this set corresponding to systems in ${\rm P_1}$ configuration (red line, 309 simulations) and $\sim46\%$ in ${\rm P_2}$ configuration (blue line, 261 simulations).

As a rule, we obtain a predominance of systems with moonlets distributed on each side of the moon than system with moonlets clustered near $L_4$/$L_5$ point. For example, for 1+4 co-orbital satellites, configurations with three moonlets near $L_4$ and one near $L_5$ are more common than configurations with all moonlets near $L_4$ but less common than the case with two moonlets near $L_4$ and two near $L_5$.

In our case of interest, we need at least three moonlets near $L_4$/$L_5$ to azimuthally confine the four arcs of Neptune. This condition is met by ${\rm P_1}$ configuration for the 1+3 co-orbital system, and ${\rm P_1}$ and ${\rm P_2}$ configurations for the 1+4 co-orbital system. These cases correspond to about 31\% of our numerical simulations, and therefore, our representative case has approximately one in three chances of producing a system of moonlets that can confine Neptune arcs.

Given this, we performed new numerical simulations by varying the mass and number of fragments to verify the robustness of our statistics. Given the simplicity of our numerical simulations, we cannot make strong claims about the formation of 1+N co-orbital satellite systems. However, with the new simulations, we intend to analyse, at least in a first approximation \citep{Tr15}, how common is the formation of systems capable of confining the arcs. The results of the new simulations are presented below.

\subsection{Varying the mass of fragments}

To evaluate the effect of the relative mass of the fragments on the results, we performed simulations assuming the following sets of fragments: \textit{m\_i)} a set varying by 50\% the mass of the fragments in relation to the representative case -- two fragments with mass $m_{\rm tro}/8$ and two with mass $3m_{\rm tro}/8$ -- and \textit{m\_ii)} a set varying the mass of the fragments by 25\% -- two pairs of fragments with masses $3m_{\rm tro}/16$ and $5m_{\rm tro}/16$. Note that the total mass in fragments is $m_{\rm tro}$ in all simulations. Fragments with the same mass are initially placed at opposite vertices of the polygon in order to conserve the linear momentum after disruption. The length of the polygon is $l=2R^>_{\rm fra}+100$~m, where $R^>_{\rm fra}$ is the radius of the largest fragment. We performed 300 numerical simulations for each set of fragments.

\begin{figure}
\centering
\includegraphics[width=\columnwidth]{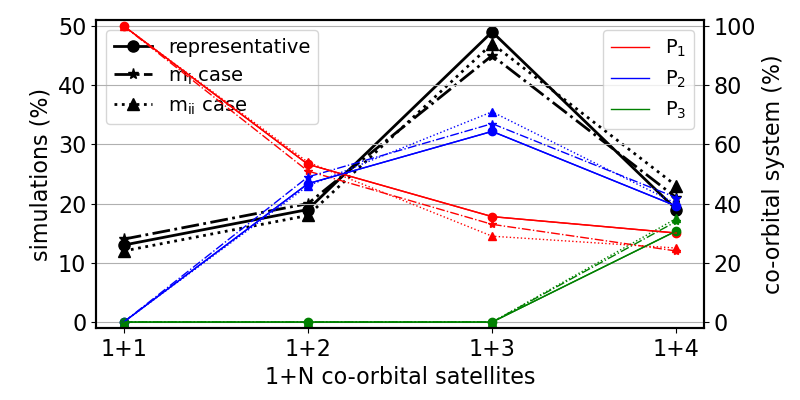}
\caption{Fractions of 1+$N$ co-orbital satellite system and {$\rm P_i$} configurations for initial fragments with masses $m_{\rm frac}=m_{\rm tro}/4$, $m_{\rm tro}/4$, $m_{\rm tro}/4$, $m_{\rm tro}/4$ (representative case, solid line with dots), $m_{\rm frac}=m_{\rm tro}/8$, $m_{\rm tro}/8$, $3m_{\rm tro}/8$, $3m_{\rm tro}/8$ (\textit{m\_i} case, dashed line with stars), and $m_{\rm frac}=3m_{\rm tro}/16$, $3m_{\rm tro}/16$, $5m_{\rm tro}/16$, $5m_{\rm tro}/16$ (\textit{m\_ii} case, dotted line with triangles). The black lines give the fraction of systems with 1+$N$ co-orbital satellites at the end of simulations, with values given on the left y-axis. The coloured lines show the fraction of simulations in {$\rm P_i$} configuration for each 1+$N$ co-orbital satellite system set, these values being given on the right y-axis. \label{esrateM}}
\end{figure}

Figure~\ref{esrateM} provides the fractions obtained for the representative case (solid line with dots) and the cases \textit{m\_i} (dashed line with stars) and \textit{m\_ii} (dotted line with triangles). The values relative to the black lines are those given on the left y-axis and correspond to the fraction of simulations that result in 1+N co-orbital satellites systems. The coloured lines correspond to the fractions of ${\rm P_i}$ configurations obtained in the systems with 1+$N$ co-orbital satellites, with values given on the right y-axis.

The fractions for the cases \textit{m\_i} and \textit{m\_ii} show small variations in relation to the representative case, indicating that variations in the mass of fragments have a small effect on the evolution of the system. The orbital evolution of the fragments is mainly defined by the azimuthal confinement due to $S_1$. As we are assuming fragments with masses two orders of magnitude smaller than the moon, we have that such a result is somehow expected.

Assuming different masses for the fragments, we obtain moonlets with a greater variety of mass in relation to the representative case. Consequently, the equilibrium positions will not be strictly the same in all systems with the same ${\rm P_i}$ configuration. However, we find that the azimuthal differences with respect to the cases shown in Figures~\ref{ptos2p}-\ref{ptos5p} are less than 1~deg for all simulations. Next, we vary the initial number of fragments.

\subsection{Varying the number of fragments}

\begin{figure}
\centering
\includegraphics[width=\columnwidth]{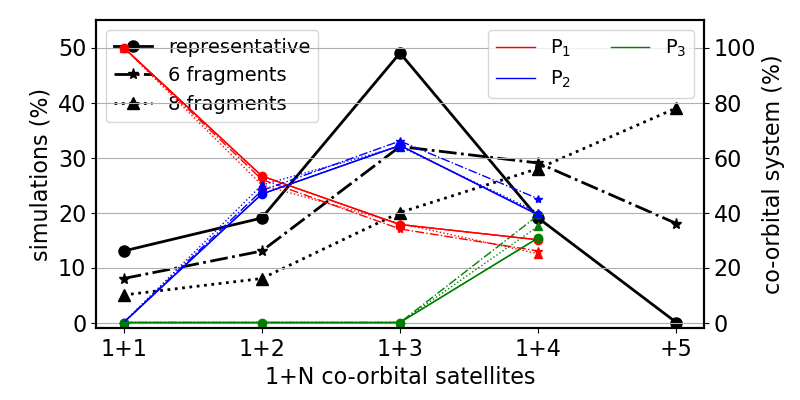}
\caption{Fractions of 1+$N$ co-orbital satellite system and {$\rm P_i$} configurations for cases with 4 (representative case, solid line with dots), 6 (dashed line with stars) and 8 fragments (dotted line with triangles). The coloured lines are the fraction of systems in ${\rm P_i}$ configuration, for $N=$ 1, 2, 3, and 4 (right y-axis). The black lines give the fraction of simulations that produce $N$ moonlets (left y-axis). Position ``+5'' on x-axis corresponds to systems with 5 moonlets or more. \label{esrateN}}
\end{figure}

We performed sets of 300 numerical simulations with $N_{\rm fra}$ fragments of the same mass $m_{\rm tro}/N_{\rm fra}$. The fragments are distributed in a regular polygon with $N_{\rm fra}$ vertices and length $l=2R+100$~m. In Figure~\ref{esrateN}, we show the same fractions as in Figure~\ref{esrateM} for the representative case (solid line with dots) and for the cases with 6 (dashed line with stars) and 8 fragments (dotted line with triangles).

As a rule, cases with $N_{\rm fra}$ fragments give rise to systems with up to $N_{\rm fra}$ moonlets, and we obtain that the main effect of the number of fragments is to change the fractions of systems with 1+$N$ co-orbital satellites. The fraction of systems in ${\rm P_i}$ configuration for $N\leq 4$ is almost the same for all cases. In the case with $N_{\rm fra}=6$, 18\% of the simulations result in systems with 5 or 6 moonlets, reducing the fraction of systems with 3 or 4 moonlets compared to the representative case. However, systems with 1+3 co-orbital satellites remain the most common outcome of the simulations. For $N_{\rm fra}=8$, 39\% of the systems have more than 5 co-orbital satellites.

In section~\ref{lersys}, we obtained different equilibrium configurations with 1+3 and 1+4 co-orbital satellites that can confine the Neptune arcs. The characteristic that defines whether a configuration can confine four arcs is the presence of at least three moonlets near to the moon's $L_4$/$L_5$, which is met for all equilibrium configurations with more than 5 co-orbital satellites. Although the fraction of systems with 3 and 4 moonlets decreases when the number of fragments increases, the fractions of systems that can confine the four arcs are greater than that obtained in our representative case (31\%). Therefore, the case with four fragments can be interpreted as a lower bound in our analysis of how common is the formation of a system capable of confining the arcs. In the next section, we cover the production of debris in the system.

\section{Comments on arc formation} \label{arcsloc}

The origin of Neptune arcs remains a topic of debate among planetary scientists. In light of the confinement model of \cite{Po91}, it was proposed that the arcs would originate from dust ejected by immersed satellites in resonances with Galatea. However, we now know that this confinement model is not applicable to the arcs \citep{Si99,Du02}, being required a different origin for them.

Several works propose that collisions between macroscopic particles are the source of the dust content observed in the Neptune arcs \citep{Sm89,Co90,Po95,salo1998}. It is an attractive proposition given the observational evidence for the existence of metric bodies immersed in the arc Fraternit\'e \citep{Pa05}. In the context of their confinement model, \cite{Re14} propose a hierarchical scenario where a previously accreted satellite gathers material at its Lagrangian point, forming the moonlets and the arcs. Furthermore, the hypothesis that the structures originate from the breakup of a parent satellite is mentioned in \cite{Pa19}.

Generally speaking, the Neptune arcs are likely composed of material produced by different processes. In the context of the scenario proposed in Section~\ref{4frag}, we envision the production of material in three different stages of the system: At the disruption of trojan moon, in the later stage, in which the fragments are evolving to form the moonlets and after the formation of the 1+N co-orbital satellite system. Below, we discuss the mechanisms involved in each stage. We need to emphasize that collisions are complex events that are not yet well understood, and the formation of Neptune arcs requires a careful study that is beyond the scope of our work. We do not intend here to reproduce the arcs, but only to comment on possible sources for them.

\subsection{Moon disruption stage}
The disruption of objects with a few kilometres in radius due to a $\sim$km/s impact simultaneously produces a set of larger fragments and a large amount of debris with sizes ranging from micrometres to metres \citep{Be99,Mi04}. The former were assumed by us as the building blocks of the moonlets, while the latter may have contributed to the formation of the arcs. As shown in \cite{Ga20}, the dust material in the Adams ring region has a short lifetime due to solar radiation and plasma drag, so macroscopic debris are the most likely to have contributed to the formation of the arcs.
\begin{figure*}
\centering
\includegraphics[width=1.5\columnwidth]{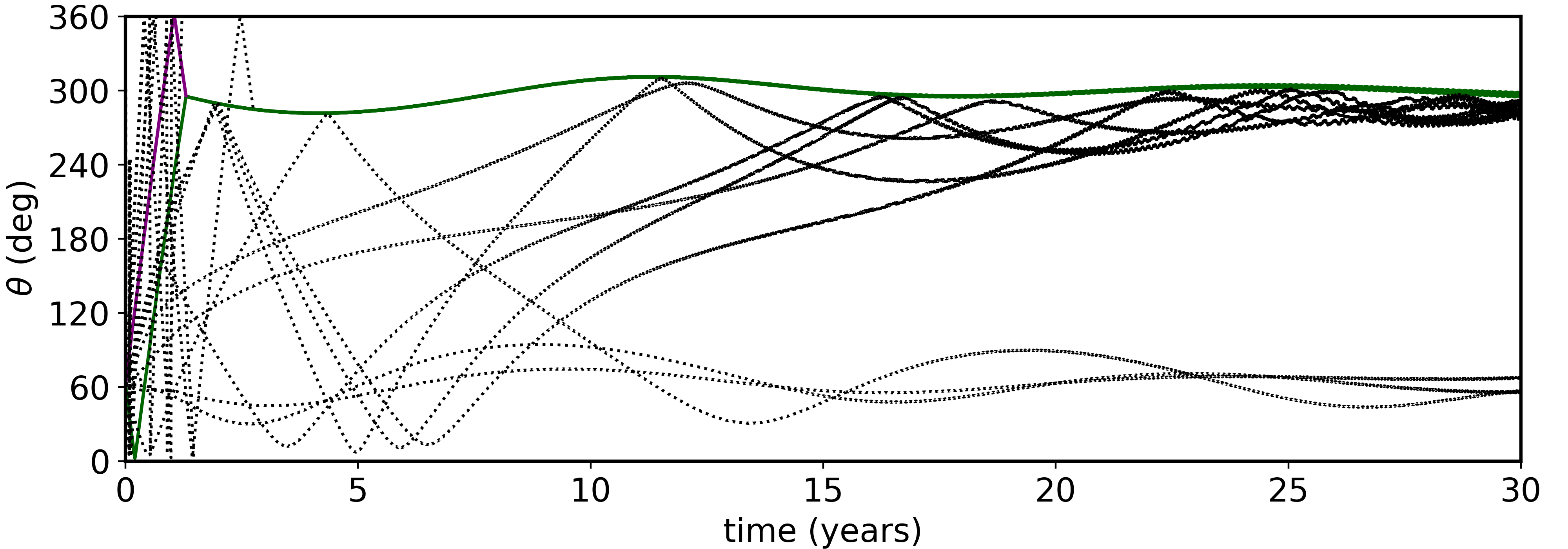}
\caption{Longitudinal evolution of four fragments (coloured lines) and a set of particles (black lines) initially distributed in the circle circumscribing the polygon of the fragments. The moonlet settles in the $L_5$ point and two arcs are formed, near $L_4$ and $L_5$ points. \label{sc1}}
\end{figure*}

To analyse the evolution of such particles, we redid some numerical simulations of the representative case, distributing 500 massless particles randomly in the circle that circumscribes the fragments' polygon, with randomly chosen radial ejection velocities of $0.36-0.73$~m/s. The non-conservative term was also applied to the particles. The simulation with the highest particle survival rate is presented in Figure~\ref{sc1}, where the fragments give rise to 1+1 co-orbital satellites in ${\rm P_1}$ configuration. After impacts in the first years of simulation, the moonlet settle in the equilibrium position in less than 10~years and particles are confined azimuthally to two of the three particle equilibrium positions of the system.

In general, we obtain a low particle survival, with more than 70\% of the set colliding with the fragments in five years of simulation. We do not verify particle survival in the cases of our interest (1+3 and 1+4 co-orbital satellite systems), whereas up to 5\% of the particles survive in the 1+1 and 1+2 co-orbital satellite systems. These results seem to indicate that the debris formed in the disruption does not directly contribute to the arcs, or only contributes with a small amount of material. However, the relatively high impact velocities (${\rm \sim m/s}$) indicate erosive events \citep{St12}. Therefore, they should give rise to a second generation of debris that contribute to the arcs, as will be discussed ahead. 

\subsection{Moonlets formation stage}

Azimuthal confinement due to $S_1$ and Galatea's gravitational effect increase collisions between debris, which can be a significant source of material to the arcs. Just as a proof of concept, we assume the total debris mass as $m_{\rm tro}/4$, following a distribution given by $N\propto s^{-2.4}$ \citep{Kr03} (${\rm s=1~\mu m-1~m}$) and we calculate the rate of mass-produced due to impacts between debris. This can be estimated for soft target-ejecta as \citep{Co90}
\begin{equation}
\dot{M}_{\rm coll}=4\times10^{-8}\frac{\Omega}{A}\sum_{s_i>s_j}N_i\sum_{s_j}N_jK_j(s_i+s_j)^2  \label{mcoll}  
\end{equation}
where $\Omega$ is the orbital frequency, $A$ is the area of the region, and $N_{i,j}$ is the number of debris with radius $s_{i,j}$. $K_j$ is the kinetic energy of the impactor particle, where we have assumed impacts with mean velocities of ${\rm v_{\rm imp}=1~m/s}$.

As a result, we obtain that impacts can populate the four arcs in $T\sim 10^4$~years (optical depth of $\tau=0.1$), showing that such events can be the source of the arcs. Using $\dot{M}_{\rm coll}$, we made a rough estimate comparing the cross-section of the debris with that of the fragment and found that the production, due to debris-fragment impacts, can reduce the time $T$ by one order of magnitude.

As we showed in Section~\ref{4frag}, some impacts between fragments are likely to be events with partial merging or erosion and therefore will also give rise to a second generation of debris that can contribute to the arcs. Based on \cite{Ca95}, \cite{Su15} assume that 12\% of a moon's regolith layer is released in a collision between two moonlets. As an estimate only, we made the humblest assumption that 1\% of a fragment's mass is released in a collision between two fragments, giving an output of $6\times 10^{10}$~kg per collision. This amount of material exceeds the estimated mass of the arcs ($4\times 10^{10}$~kg). Therefore, if completely confined, the material produced in just one impact between fragments is sufficient to reproduce the optical depth of the four arcs.

In order to analyse the evolution of the second generation debris, we redid some numerical simulations, distributing 500 massless particles in a disk around the moonlet formed right after a collision. The particles are influenced by the non-conservative term, and the distance of each particle to the moonlet is chosen randomly in the range ${\rm 1.0-1.5~R_b}$, where ${\rm R_b}$ is the physical radius of the body. The radial velocity is chosen randomly in the range ${\rm 1-3~v_{esc}}$. For higher velocities, particles leave the horseshoe region. 

Most particles collide in the first few years after being added to the system. However, the fraction that eventually survives is greater than in the case of particles produced in the disruption. This was to be expected, as particles are added later in the system, sometimes when satellites are already in tadpole-like orbits. In 15\% of the numerical simulations in ${\rm P_1}$ configuration with 1+3 co-orbital satellites, at lest one arc with material is obtained at the end of the simulation.

\begin{figure*}
\centering
\includegraphics[width=1.5\columnwidth]{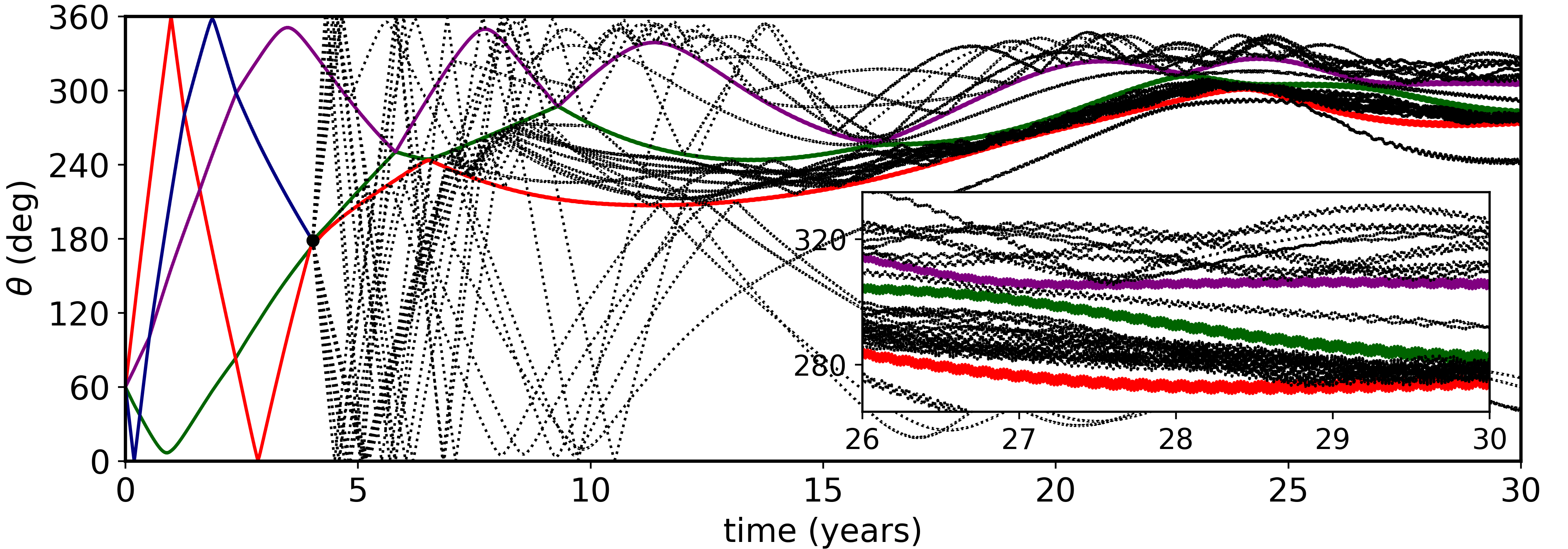}
\caption{Evolution of a set of particles produced by the collision of two fragments in a simulation initially with four fragments that give rise to a system in ${\rm P_1}$ configuration with 1+3 co-orbital satellites. The particles are initially in a disk around the moonlet formed after the collision. The instant of the collision ($\sim 4$~yr) is set by the black dot. We only show the surviving particles (black dotted lines), which are about 8\% of the initial set. Fragments that give rise to the satellites are the solid coloured lines. \label{sc2}}
\end{figure*}

Figure~\ref{sc2} shows a case where all particle equilibrium positions are populated by material at the end of simulation, with a particle survival rate of $8\%$. As can be seen, four arcs are obtained near the $L_5$ point of the moon. This case is an exception, in most simulations only two arcs were produced, but it serves our purpose to demonstrate that collisions between fragments can be the origin of at least some of the arcs.

\subsection{Post-formation stage}

After the moonlets are formed, they can suffer impacts of  interplanetary dust particles (IDPs) or meteoroids originating mainly from the Kuiper Belt \citep{Po16,Po19}. These impacts can provide material for the Adams ring and   also be part of the source of Neptune arcs. IDPs have a typical radius of order ${\rm 100~\mu m}$, while meteoroids can be up to a few metres in size. At the end of the section, we briefly discuss different sizes of material produced by these two different populations of impactors.

\begin{figure*}
\centering
\subfigure[]{\includegraphics[width=1.5\columnwidth]{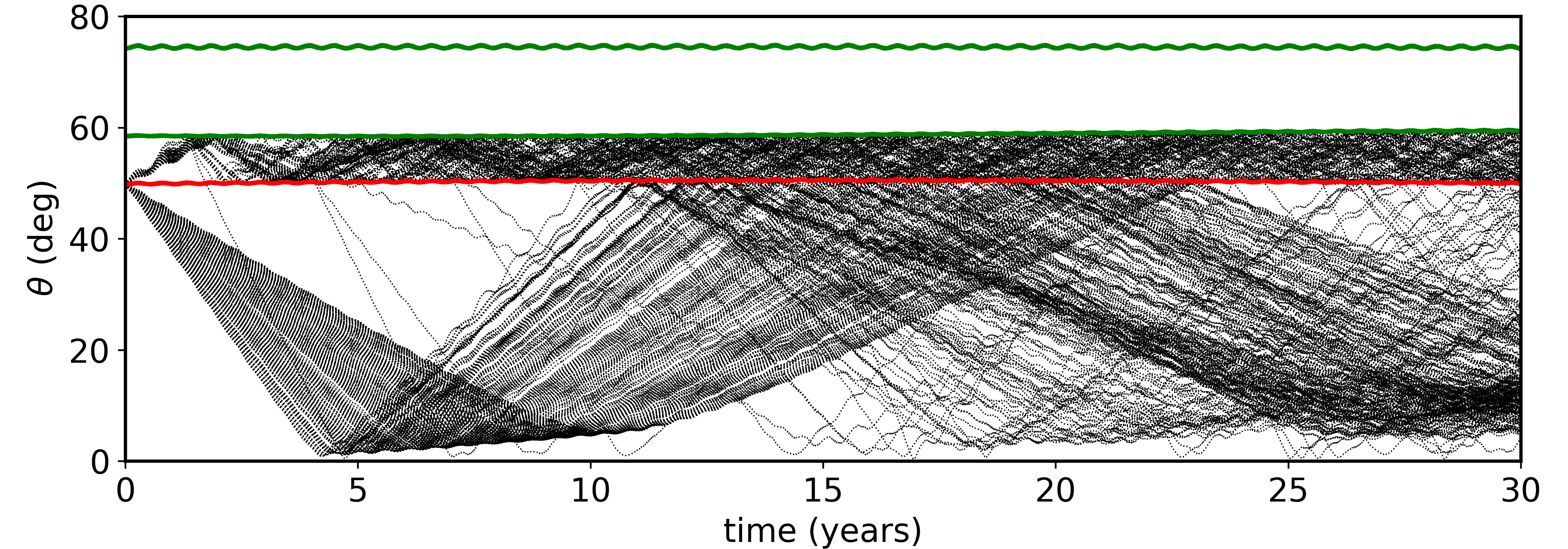} \label{sc31}}
\subfigure[]{\includegraphics[width=1.5\columnwidth]{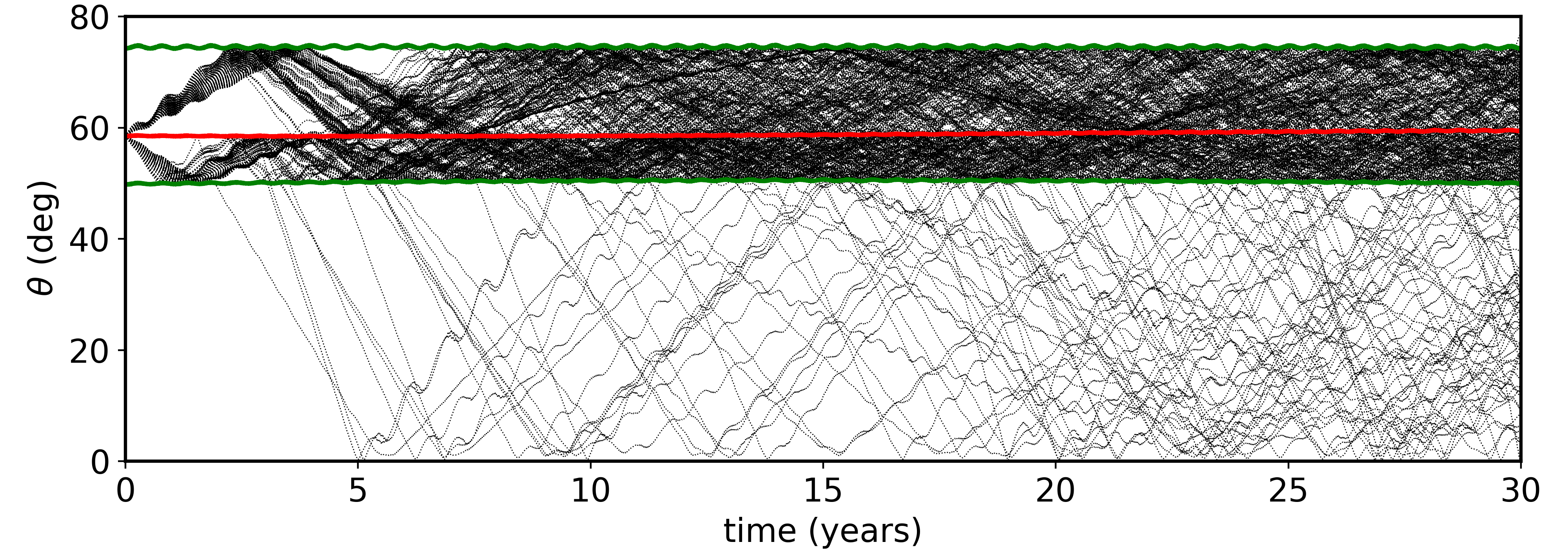} \label{sc32}}
\subfigure[]{\includegraphics[width=1.5\columnwidth]{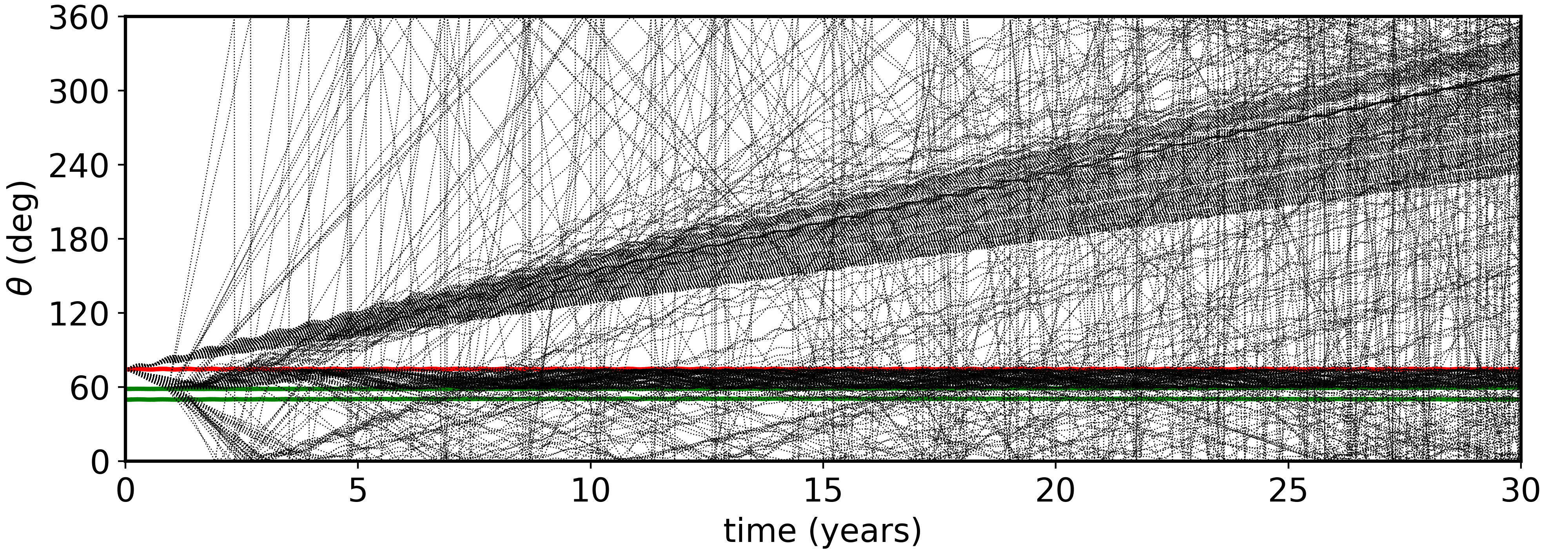} \label{sc33}}
\caption{Temporal evolution of particles produced by impacts of external bodies for a ${\rm P_1}$ configuration with 1+3 co-orbital satellites. We show in separate panels the particles (black dotted lines) that originated from different moonlets. The moonlets are shown in coloured lines. The one that produces the material is the red line, and other moonlets are the green lines. We refer to the moonlets from bottom to top as $S_2$, $S_3$, and $S_4$. \label{sc3}}
\end{figure*}

We distributed 500 particles in a disk around the moonlets. The distance of particles to the moonlet is chosen randomly in the range ${\rm 1.0-1.5~R_b}$ and their radial velocities are chosen randomly in the range ${\rm 1-3~v_{esc}}$. Figure~\ref{sc3} shows the evolution of the particles in black dotted lines for a ${\rm P_1}$ configuration with 1+3 co-orbital satellites. Each panel corresponds to a different simulation, where particles are initially around a different moonlet of the system. The moonlet which produces the particles is shown in red, while the others are in green. We found that $<10\%$ of the particles collide in all the simulations.

As expected, the particles are confined by a pair of moonlets, as we can see for example in Figure~\ref{sc2}b, where most of the particles launched by $S_3$ are confined between $S_2-S_3$ and $S_3-S_4$. A few particles leave the confinement of the pair $S_2-S_3$, becoming confined by the pair $S_1-S_2$; these particles correspond to the transient particles classified by \cite{Gi20}.

We emphasize that the type of impactor influences the sizes of the launched particles. The impacts of IDPs produce micrometre particles (${\rm 1-100~\mu m}$) that suffer the effects of dissipative forces, such as the solar radiation force and plasma drag \citep[see][]{Ga20}. \cite{Gi20} showed that the lifetime of micrometre particles in the arcs under the effects of the solar radiation force is less than 50~years and that moonlets cannot replenish the arcs by IDP impacts.

On the other hand, meteoroid collisions can produce from micrometre particles up to metric-sized debris. For centimetric or larger bodies, the effects of the dissipative forces can be disregarded and the orbital evolution of the bodies is represented by Figure~\ref{sc3}. \cite{Gi20} showed that larger bodies survive for more than 1,000~years in the arcs, making meteoroid impacts a possible source for the arcs. The arcs' formation due to meteoroid impacts is an intricate problem, as collisions cannot disturb the stability of co-orbital satellites while they must produce an amount of material that reproduces the system.

\section{Discussion} \label{discussion}

In this work, we analyse through a set of numerical simulations the formation and orbital evolution of 1+$N$ co-orbital satellite systems that could confine the arcs of Neptune. Revisiting the work of \cite{Re04}, we obtain that the equilibrium configurations obtained by them are not altered when we consider the moonlets in an orbit with the estimated eccentricity of particles located at the Adams ring \citep{Re14}. It turns out that the 42:43 LER with Galatea does not destroy the equilibrium configurations, but only shifts the equilibrium positions by a few degrees.

We obtained a total of three distinct equilibrium configurations that can reproduce the angular width of the arcs -- ${\rm P_1}$ configuration with 1+3 co-orbital satellites and the ${\rm P_1}$ and ${\rm P_2}$ configurations with 1+4 co-orbital satellites. We also have that small variations in the mass of the moonlets do not alter the equilibrium configurations \citep{Re14}. These results are interesting, as they demonstrate that different set of masses, longitudes, and number of moonlets can confine the arcs, giving robustness to the model proposed by \cite{Re14}. 

The origin of these moonlets and arcs is still unknown, being discussed in \cite{Re14} a scenario in which an ancient satellite gathered material around its $L_4/L_5$ point, forming the moonlets. The arcs would be the residual material of this process. Although mechanisms such as the one proposed by \cite{Iz10} show the formation of moons at the triangular points of satellites, it is not clear which process would lead to the formation of some moonlets and not just one. It is also unclear how these moonlets would reach the equilibrium configuration. Satellite formation simulations in circumplanetary disks show the formation of pairs of co-orbitals \citep{Ma21}, but not of systems with more than two co-orbitals. Such results indicate that the moonlets, if they exist, probably have formed by a process other than simple accretion.

We propose the formation of moonlets by the disruption of an ancient body due to an impact with an ongoing object. This is a possible scenario since many objects originating in the Kuiper belt crosses the Neptune region \citep{Co92,Le00}. We are aware that our treatment is very simplistic. A more realistic model requires the study of the disruption itself, which depends on the physical parameters of target and impactor, impact parameter and velocity. It also requires studying the post-evolution of fragments and debris, considering mechanisms such as growth due to collisions with smaller debris and Adams ring material.

Nevertheless, our simulations varying the mass of fragments result in variation in the fractions lower than $10\%$, showing a self-consistency. In these simulations, we obtain that a disruption has a probability of $\sim 30\%$ of producing a system capable of confining the four arcs. When we assume the formation of more than four fragments in the disruption, these values increase due to the possible formation of a system with more than 5 moonlets. These results place the scenario studied here, at least in a first approximation, as a possible scenario for the formation of the moonlets.

Our simulations include an artificial non-conservative term responsible for varying the system's energy and carrying the moonlets to equilibrium positions. In our representative case, this term was adjusted to generate the equilibrium configuration over an arbitrary period of 30 years, and we also performed simulations with timespans of 300 and 3000~years. In the real system, a series of mechanisms act by varying the energy of the system, such as inelastic collisions with disruption debris and the Adams ring material, and resonant torques due to Galatea. However, such effects work over long timescales, indicating that the moonlets formed and settled at equilibrium positions over timescales longer than the ones considered by us.

Changes in brightness and longitude of the arcs \citep{Pa05} and the disappearance of the arcs Libert\'e and Courage \citep{Pa19} seem to indicate that the dust population of the arcs is recent. \cite{Gi20} support this assumption by obtaining lifetimes of a few decades for the dust material in the arcs. \citeauthor{Gi20} also found that the disappearance of two arcs can be explained if they are composed of particles with typical sizes different from the arcs that still remain.

In our scenario, these differences in the particle sizes between the arcs can be obtained if they originated at different stages. In particular, our crude analysis shows that micrometre-sized material is possibly originated from impacts between the disruption outcomes (fragments and debris). Meteoroid impacts with the already formed moonlets is another possible source. However, it is likely that several processes act to produce the arcs, such as impacts, fragmentation, and erosion.

\section{Conclusion} \label{conclusions}
We have explored, by numerical simulations, the confinement model for Neptune arcs proposed in \cite{Re14}. The model proves to be possible, as different sets of moonlets, in number, mass, and location, can roughly reproduce the width of the four arcs. However, further investigation and refinements are needed to explain the interesting evolution shown by the arcs since their discovery. If these co-orbital satellites exist, we show that the disruption of an ancient body at a triangular point of a moon is a possible model for their formation. In such a scenario, the arcs may have been formed by different processes such as collisions, fragmentation, re-accretion, external impacts, among others. We find that impacts between fragments and debris and meteoroid impacts with the moonlets are attractive possibilities. The arcs may have been formed in different stages, with the arcs composed only of dust particles being the final stage of the arc life.

\section*{Acknowledgements}
We thank the referee for the comments that significantly improved our work. The authors thank FAPESP (2016/24561-0, 2018/23568-6), CNPq (313043/2020-5) and Capes (Finance Code 001) for the financial support. This research was  carried out using the computational resources of the Center for Mathematical Sciences Applied to Industry (CeMEAI) funded by FAPESP (grant 2013/07375-0).

\section*{Data Availability}
The data underlying this article will be shared on reasonable request to the corresponding author.

%
%
\bibliographystyle{mnras}
\bibliography{aanda} 




\bsp	
\label{lastpage}


%

\end{document}